\documentclass[journal,twoside]{IEEEtran}

\usepackage{amsmath}
\usepackage{amsthm_noproof}
\usepackage{amssymb}
\usepackage{bm}
\usepackage{graphicx}
\usepackage{cite}

\graphicspath{{plot/}}

\newtheorem{theorem}{Theorem}

\newtheorem{lemma}{Lemma}

\newcommand{\varn}{\ensuremath{\sigma_{n}^{2}}}
\newcommand{\rvq}{\textsf{rvq}}
\newcommand{\B}{\bar{B}}
\newcommand{\N}{\bar{N}_r}
\newcommand{\K}{\bar{K}}
\newcommand{\diff}{\mathrm{d}}
\newcommand{\mf}{\textsf{mf}}
\newcommand{\mmse}{\textsf{mmse}}
\newcommand{\var}{\ensuremath{\mathrm{var}}}
\newcommand{\tr}{\ensuremath{\mathrm{tr}}}

\newcommand{\me}{\ensuremath{\mathrm{e}}}

\newcommand{\bH}{\ensuremath{\bm{H}}}
\newcommand{\srvq}{\gamma_{\rvq}^{\infty}}
\newcommand{\bLd}{\ensuremath{\bm{\Lambda}}}
\newcommand{\bU}{\ensuremath{\bm{U}}}
\newcommand{\bw}{\ensuremath{\bm{w}}}

\newcommand{\lmax}{\ensuremath{\lambda_{\max}^{\infty}}}
\newcommand{\bv}{\ensuremath{\bm{v}}}

\newcommand{\bV}{\ensuremath{\bm{V}}}

\begin{document}

\title{Capacity of a Multiple-Antenna Fading Channel With a Quantized
  Precoding Matrix}

\author{Wiroonsak~Santipach,~\IEEEmembership{Member,~IEEE,} and
Michael L. Honig,~\IEEEmembership{Fellow,~IEEE}%
\thanks{Manuscript recieved December 22, 2006; revised July 17, 2008.
This work was supported by the U.S. Army Research Office under grant
DAAD19-99-1-0288 and the National Science Foundation under grant
CCR-0310809, and was presented in part at IEEE Military Communications
(MILCOM), Boston, MA, USA, October 2003, IEEE International Symposium
on Information Theory (ISIT), Chicago, IL, USA, June 2004, and IEEE
International Symposium on Spread Spectrum Techniques and Applications
(ISSSTA), Sydney, Australia, August 2004.}%
\thanks{W. Santipach was with the Department of Electrical Engineering
and Computer Science; Northwestern University, Evanston, IL 60208 USA.
He is currently with the Department of Electrical Engineering; Faculty
of Engineering; Kasetsart University, Bangkok, 10900 Thailand (email:
wiroonsak.s@ku.ac.th).}%
\thanks{M. L. Honig is with the Department of Electrical Engineering
and Computer Science; Northwestern University, Evanston, IL 60208 USA
(email: mh@eecs.northwestern.edu).}%
\thanks{Communicated by H. Boche, Associate Editor for
Communications.}%
\thanks{Digital Object Identifier 10.1109/TIT.2008.2011437}}

\markboth{IEEE TRANSACTIONS ON INFORMATION THEORY, VOL. 55, NO. 3, MARCH
2009}{SANTIPACH AND HONIG: CAPACITY OF A MULTIPLE-ANTENNA FADING CHANNEL
WITH A QUANTIZED PRECODING MATRIX}

\IEEEpubid{0018--9448/\$25.00~\copyright~2009 IEEE}

\maketitle

\begin{abstract}
Given a multiple-input multiple-output (MIMO) channel, feedback from
the receiver can be used to specify a transmit precoding matrix, which
selectively activates the strongest channel modes.  Here we analyze
the performance of {\em Random Vector Quantization (RVQ)}, in which
the precoding matrix is selected from a random codebook containing
independent, isotropically distributed entries.  We assume that
channel elements are {\em i.i.d.}  and known to the receiver, which
relays the optimal (rate-maximizing) precoder codebook index to the
transmitter using $B$ bits. We first derive the large system capacity
of beamforming (rank-one precoding matrix) as a function of $B$, where
large system refers to the limit as $B$ and the number of transmit and
receive antennas all go to infinity with fixed ratios.  RVQ for
beamforming is asymptotically optimal, i.e., no other quantization
scheme can achieve a larger asymptotic rate.  We subsequently consider
a precoding matrix with arbitrary rank, and approximate the asymptotic
RVQ performance with optimal and linear receivers (matched filter and
Minimum Mean Squared Error (MMSE)).  Numerical examples show that
these approximations accurately predict the performance of finite-size
systems of interest.  Given a target spectral efficiency, numerical
examples show that the amount of feedback required by the linear MMSE
receiver is only slightly more than that required by the optimal
receiver, whereas the matched filter can require significantly more
feedback.
\end{abstract}


%%%%%%%%%%%%%%%%%%%%%%%%
\section{Introduction}
%%%%%%%%%%%%%%%%%%%%%%%%

\IEEEPARstart{G}{iven} a multi-input multi-output (MIMO) channel,
providing channel information at the transmitter can increase the
achievable rate and simplify the coder and decoder.  Namely, this
channel information can specify a precoding matrix, which aligns the
transmitted signal along the strongest channel modes (i.e., singular
vectors corresponding to the largest singular values).  In practice,
the precoding matrix must be quantized at the receiver, and relayed to
the transmitter via a feedback channel. The corresponding achievable
rate is therefore limited by the accuracy of the quantizer.

The design and performance of quantized precoding matrices for
multi-input single-output (MISO) and MIMO channels has been considered
in numerous references, including
\cite{narula98,mukkavilli03,love03,LoveHeath05,lau04,roh_it06,
roh_sp06,dai_liu,dai_liu_rider_Grassman,zhou05,xia06,mukkavilli03asilomar}.
In those references, and in this paper, the channel is assumed to be
stationary, known at the receiver, and the performance is evaluated as
a function of the number of quantization bits $B$. (This is in
contrast with other work, which models estimation error at the
receiver, but does not explicitly account for quantization error
(e.g., \cite{jongren02,skoglund03}), and which assumes a time-varying
channel with feedback of second-order statistics
\cite{visotsky01,zhou02,simon03,jafar04}.)  Optimization of vector
quantization codebooks is discussed in
\cite{narula98,love03,mukkavilli03,lau04,roh_it06} for beamforming,
and in \cite{LoveHeath05,mukkavilli03asilomar,roh_sp06} for MIMO
channels with precoding matrices that provide multiplexing gain (i.e.,
have rank larger than one).  It is shown in \cite{love03,mukkavilli03}
that this optimization can be interpreted as maximizing the minimum
distance between points in a Grassmannian space. (See also
\cite{dai_liu_rider_Grassman}.)  The performance of this class of
Grassmannian codebooks is also studied in
\cite{zhou05,dai_liu,dai_liu_rider_Grassman}.

\IEEEpubidadjcol In this paper, we evaluate the performance of a {\em
Random Vector Quantization (RVQ)} scheme for the precoding
matrix. Namely, given $B$ feedback bits, the precoding matrix is
selected from a random codebook containing $2^B$ matrices, which are
independent and isotropically distributed. RVQ has been analyzed in
other source coding contexts (e.g., see \cite{Zador} and the related
discussion in \cite{GrayNeuhoff}), and achieves the rate-distortion
bound for ergodic Gaussian sources.  This work is motivated by prior
work \cite{cdma04} in which RVQ is considered for signature
quantization in Code-Division Multiple Access (CDMA).  In that
scenario, limited feedback is used to select a signature for a
particular user, which maximizes the received
Signal-to-Interference-Plus-Noise-Ratio (SINR).  RVQ has the
attractive properties of being tractable and asymptotically
optimal. Namely, in \cite{cdma04} the received SINR with RVQ is
evaluated in the asymptotic (large system) limit as processing gain,
number of users, and feedback bits all tend to infinity with fixed
ratios. Furthermore, it is shown that no other quantization scheme can
achieve a larger asymptotic SINR.

Here we assume an {\em i.i.d.} block Rayleigh fading channel model
with independent channel gains, and take ergodic capacity as the
performance criterion.  The receiver relays $B$ bits to the
transmitter (per codeword) via a reliable feedback channel (i.e., no
feedback errors) with no delay. We start by evaluating the capacity of
MISO and MIMO channels with a quantized {\em beamformer}, i.e.,
rank-one precoding matrix.  Our results are asymptotic as the number
of transmit antennas $N_t$ and feedback bits $B$ both tend to infinity
with fixed $B/N_t$ (feedback bits per degree of freedom).  For the
MIMO channel the number of receive antennas $N_r$ also tends to
infinity in proportion with $N_t$ and $B$.  The asymptotic expressions
accurately predict the performance of finite-size systems of interest
as a function of normalized feedback and background Signal-to-Noise
Ratio (SNR).  In analogy with the optimality result shown in
\cite{cdma04}, RVQ is also asymptotically optimal in this scenario,
i.e., no other quantization scheme can achieve a larger asymptotic
rate. Furthermore, numerical examples for small $N_t$ show that RVQ
performance averaged over codebooks is essentially the same as that
obtained from codebooks optimized via the Lloyd-Max algorithm
\cite{narula98,lau04,roh_it06}.  (See also the numerical examples in
\cite{commag04}, which compare RVQ performance with the optimized
(Grassmannian) codebooks in \cite{love03}.)

We then consider quantization of a precoding matrix with arbitrary
rank. Namely, a rank $K$ precoding matrix multiplexes $K$ independent
streams of transmitted information symbols onto the $N_t$ transmit
antennas.  In that case, the capacity with limited feedback is
approximated in the limit as $B$, $N_t$, $N_r$, and $K$ all tend to
infinity with fixed ratios $K/N_t$, $N_r/N_t$, and $B/N_r^2$.  That
is, the number of feedback bits again scales linearly with the number
of degrees of freedom, which is proportional to $N_r^2$. Although our
results for beamforming suggest that RVQ is also asymptotically
optimal in this scenario, this remains an open question.

The asymptotic results for a precoder matrix with arbitrary rank $K$
can be used to determine the normalized rank, or multiplexing gain
$K/N_t$, which maximizes the capacity.  This optimized rank in general
depends on the normalized feedback, the ratio of antennas $N_t/N_r$,
and the SNR.  For example, if $N_t/N_r \geq 1$ and the SNR is
sufficiently large, then as the feedback increases from zero to
infinity, the optimized rank decreases from one to $N_r/N_t$.
Numerical results are presented, which illustrate the effect of
normalized rank on achievable rate, and also show that the asymptotic
results accurately predict simulated results for finite-size systems
of interest.

We also evaluate the performance of RVQ with linear receivers (i.e.,
the matched filter and linear Minimum Mean Squared Error (MMSE)
receivers), and compare their performance with the optimal
(capacity-achieving) receiver.  With the optimal precoding matrix,
corresponding to infinite feedback, both linear receivers are optimal.
With limited feedback the two linear receivers are simpler than the
optimal receiver, but require more feedback to achieve a target
rate. Numerical results show that this additional feedback required by
the linear MMSE receiver is quite small, whereas the additional
feedback required by the matched filter can be significant (e.g.,
about one bit per precoding matrix element).

In addition to quantizing the optimal precoding matrix, power for each
data stream can also be optimized, quantized, and fed back to the
transmitter (e.g., see \cite{bhashyam02,lau02}).  Asymptotically, the
amount of feedback required to specify the power is negligible
compared to the feedback required for the precoding matrix.
Furthermore, uniform power over the set of activated channel typically
performs close to the optimal (water-filling) performance
\cite{dai_liu}.  We therefore only consider quantization of the
precoding matrix.

Other related work on RVQ for MIMO channels has been presented in
\cite{chun_love,jindal,goldsmith_jindal_isit}.  Namely, exact
expressions for the ergodic capacity with beamforming and RVQ for a
finite-size MISO channel are derived in \cite{chun_love}.  The
performance of RVQ for precoding over a broadcast MIMO channel is
analyzed in \cite{jindal,goldsmith_jindal_isit}.  A closely related
random beamforming scheme for the multiuser MIMO broadcast channel was
previously presented in \cite{sharif}. In that work, the growth in sum
capacity is characterized asymptotically as the number of users
becomes large with a {\em fixed} number of antennas.  (Random
beamforming was previously proposed in \cite{viswanath_oppo_ant},
although there the main focus is to improve fairness among users.)

The paper is organized as follows.  Section~\ref{chnnl_mod} describes
the channel model, Section~\ref{beam} considers the capacity of
beamforming with limited feedback, and sections~\ref{opt} and
\ref{linear} examine the capacity of a quantized precoding matrix with
optimal and linear receivers, respectively.  Derivations of the main
results are given in the appendices.

%%%%%%%%%%%%%%%%%%%%%%%%%
\section{Channel Model}
\label{chnnl_mod}
%%%%%%%%%%%%%%%%%%%%%%%%%

We consider a point-to-point, flat Rayleigh fading channel with $N_t$
transmit antennas and $N_r$ receive antennas.  Let $\bm{x} = [ x_k ]$
be a $K \times 1$ vector of transmitted symbols with covariance matrix
$\bm{I}_K$, where $\bm{I}_K$ is the $K \times K$ identity matrix, and
$K$ is the number of independent data streams.  The received $N_r
\times 1$ vector is given by
\begin{equation}
  \bm{y} = \frac{1}{\sqrt{K}} \bm{H} \bV \bm{x} + \bm{n}
\end{equation}
where $\bm{H} = [ h_{n_r,n_t} ]$ is an $N_r \times N_t$ channel
matrix, $\bV = [\bm{v}_1 \ \bm{v}_{2} \ \ldots \ \bm{v}_{K} ]$ is an
$N_t \times K$ precoding matrix, and $\bm{n}$ is a complex Gaussian
noise $N_r \times 1$ vector with covariance matrix $\varn
\bm{I}_{N_r}$.  Assuming rich scattering and Rayleigh fading, the
elements of $\bm{H}$ are independent, and the channel coefficient
between the $n_t$th transmit antenna and the $n_r$th receive antenna,
$h_{n_r,n_t}$, is a circularly symmetric complex Gaussian random
variable with zero mean and unit variance ($E [ | h_{n_r,n_t} |^2 ] =
1$).

We assume {\em i.i.d.} block fading, i.e., the channel is static
within a fading block, and the channels across blocks are independent.
The ergodic capacity is achieved by coding the transmitted symbols
across an infinitely large number of fading blocks. With perfect
channel knowledge at the receiver and a given precoding matrix $\bV$,
the ergodic capacity is the mutual information between $ \bm{x} $ and
$ \bm{y} $ with a complex Gaussian distributed input, averaged over
the channel, given by
\begin{equation} 
  I( \bm{x} ; \bm{y} ) = E_{\bm{H}} \left[ \log \det \left( \bm{I} +
    \frac{\rho}{K} \bm{H} \bV \bV^{\dag} \bm{H}^{\dag} \right)
    \right]
\label{eq:mut}
\end{equation}
where $ \rho = 1/\sigma_n^2 $ is the background SNR.  We wish to
specify the precoding matrix $ \bV $ that maximizes the mutual
information, subject to a power constraint $ \| \bm{v}_{k} \| \le 1$,
for $1 \le k \le K$.

With unlimited feedback, the columns of the
optimal precoding matrix, which maximizes
\eqref{eq:mut}, are eigenvectors of the channel covariance
matrix $\bm{H}^{\dag} \bm{H}$. With $B$ feedback bits per fading
block, we can specify the precoding matrix from a quantization set or
codebook $\mathcal{V} = \{ \bV_1, \cdots, \bV_{2^B} \}$ known
{\em a priori} to both the transmitter and receiver.  The receiver
chooses the $\bV_j$ that maximizes the sum mutual information, and
relays the corresponding index back to the transmitter. 
Of course, the performance (ergodic capacity) depends on
the codebook $\mathcal{V}$.

%%%%%%%%%%%%%%%%%%%%%%%%%%%%%%%%%%%%%%%%%%%%%
\section{Beamforming with Limited Feedback}
\label{beam}
%%%%%%%%%%%%%%%%%%%%%%%%%%%%%%%%%%%%%%%%%%%%%

We start with a rank-one precoding matrix, corresponding to a single
data stream ($K = 1$).  In that case, the precoding matrix is
specified by an $N_t \times 1$ beamforming vector $\bm{v}$, which
ideally corresponds to the strongest channel mode.  That is, the
optimal $\bm{v}$, which maximizes the ergodic capacity in
\eqref{eq:mut}, is the eigenvector of $\bm{H}^\dag \bm{H}$
corresponding to the largest eigenvalue.  This vector is computed at
the receiver and a quantized version is relayed back to the
transmitter.

Let $\mathcal{V} = \{ \bm{v}_1, \ldots, \bm{v}_{2^B} \}$ denote the
quantization codebook for $\bm{v}$, given $B$ feedback
bits. Optimization of this codebook has been considered in
\cite{love03,mukkavilli03,roh_it06} with outage capacity and ergodic
capacity as performance metrics.  The performance of an optimized
codebook is difficult to evaluate exactly, and is approximated in
\cite{love03,mukkavilli03,roh_it06,zhou05,dai_liu_rider_Grassman,xia06}.
Here we consider RVQ in which $\bm{v}_1 , \cdots , \bm{v}_{2^B}$ are
independent, isotropically distributed random vectors, each with unit
norm.  This is motivated by the observation that given a channel
matrix $\bm{H}$ with {\em i.i.d.}  elements, the eigenvectors of
$\bm{H}^\dag \bm{H}$ are isotropically distributed \cite{tse00}, hence
the codebook entries should be uniformly distributed over the space of
beamforming vectors.

%---------------------------------------------
\subsection{MISO Channel}
%---------------------------------------------

\begin{figure}
\centering
\includegraphics[width=3.25in]{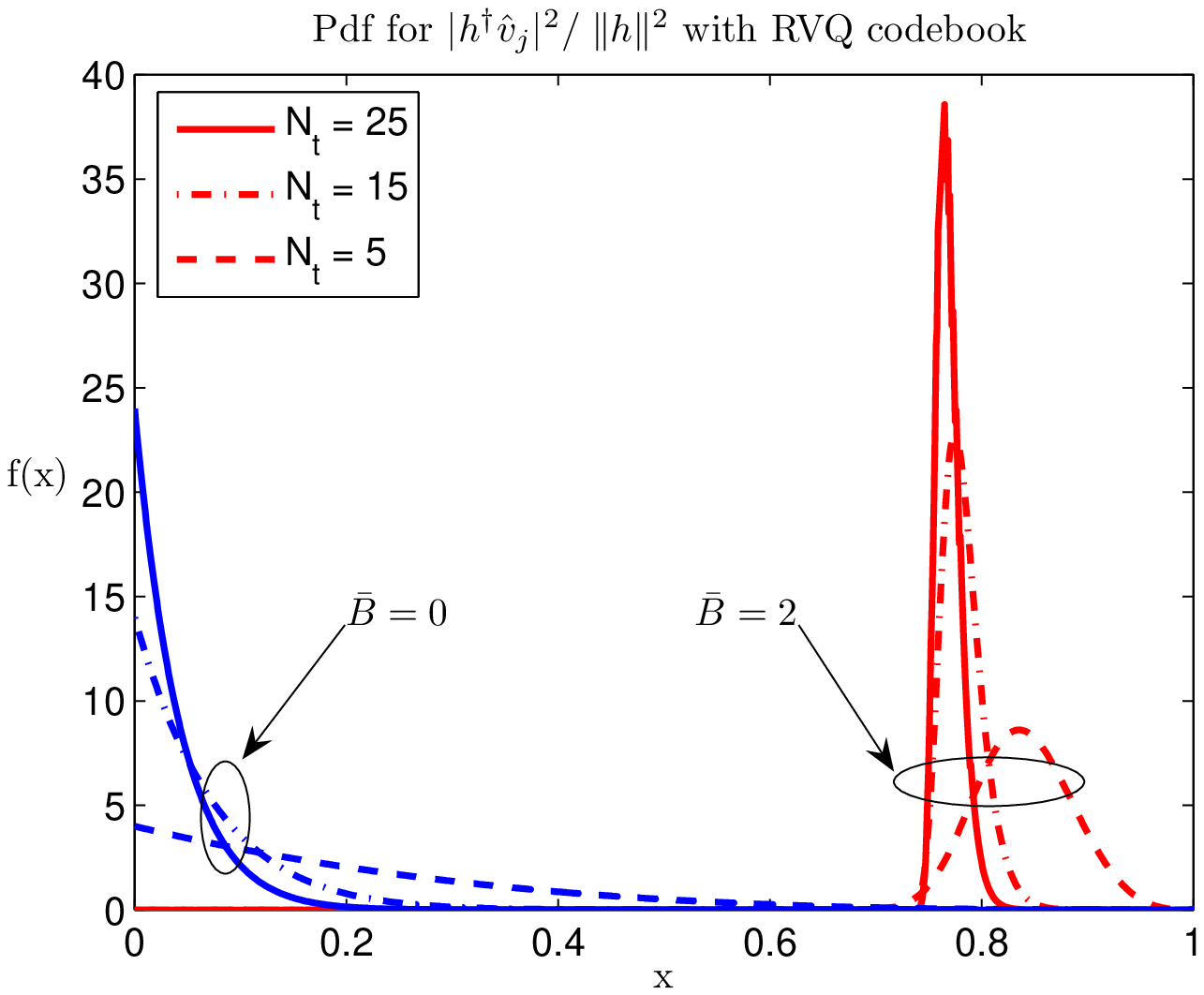}
\caption{pdf of $|\bm{h}^\dag \hat{\bm{v}}|^2 / \| \bm{h} \|^2$
  with RVQ for different values of $N_t$.}
\label{pdf_conv_miso}
\end{figure}

We first consider a MISO channel, corresponding to
a single receive antenna ($N_r = 1$).
In that case, $\bm{H}$ is an $N_t \times 1$
channel vector, which we denote as $\bm{h}$.  
The optimal beamformer, which maximizes the 
mutual information in (\ref{eq:mut}), is the normalized channel
vector $\bm{h} / \| \bm{h} \|$ and the corresponding mutual
information is
$E_{\bm{h}} [ \log ( 1 + \rho \bm{h}^{\dag} \bm{h} ) ]$.  
The receiver selects the quantized precoding vector
to maximize the mutual information, i.e.,
\begin{equation}
  \hat{\bm{v}} = \arg \max_{1 \le j \le 2^B} \left\{ 
  I_j = \log (1 + \rho | \bm{h}^{\dag} \bm{v}_j |^2 ) \right\}
\label{hbv}
\end{equation}
and the corresponding achievable rate is 
\begin{equation}
I_{\rvq}^{N_t} \triangleq \max_{1 \le j \le 2^B} I_j .
\label{eq:exp}
\end{equation}
where the superscript $N_t$ denotes the system size.
The achievable rate depends on the codebook $\mathcal{V}$
and the channel vector $\bm{h}$, and is therefore random.
Rather than averaging $I_{\rvq}^{N_t}$ over 
$\mathcal{V}$ and $\bm{h}$ to find the ergodic capacity,
we instead evaluate the limiting performance as 
$N_t$ and $B$ tend to infinity with fixed
$\bar{B} = B/N_t$ (feedback bits per transmit antenna).
In this limit, $I_{\rvq}^{N_t}$ converges to a deterministic
constant. This is illustrated in Fig. \ref{pdf_conv_miso},
which shows the pdf of $|\bm{h}^\dag \hat{\bm{v}}|^2 / \| \bm{h} \|^2$
for different $N_t$ with no feedback ($\bar{B}=0$), 
and for RVQ with $\bar{B}=2$.
The figure shows that convergence of the pdf to a
point mass is faster with feedback than without.

As $N_t \to \infty$, $(\bm{h}^\dag \bm{h})/N_t \to 1$ almost surely,
so that $\log(1+\rho\bm{h}^\dag \bm{h}) - \log(\rho N_t)\to 0$.
That is, with perfect channel knowledge at the transmitter,
the ergodic capacity increases as $\log(\rho N_t)$.
With finite feedback there is a rate loss, which is defined as
\begin{equation}
  I_{\rvq}^{\triangle} = I_{\rvq}^{N_t} - \log(\rho N_t) .
\end{equation}
For finite $N_t$, $I_{\rvq}^{\triangle}$ is random;
however, in the large system limit
$I_{\rvq}^{\triangle}$ converges to a deterministic constant.

\begin{theorem}
\label{thm:miso}
As $(N_t,B) \to \infty$ with fixed $\bar{B} = B/N_t$, the rate
difference $I_{\rvq}^{\triangle}$ converges in the 
mean square sense to
\begin{equation}
  \mathcal{I}_{\rvq}^{\triangle} = \log ( 1 - 2^{-\bar{B}} ) .
\label{eq:miso}
\end{equation}
\end{theorem}

The proof is given in Appendix~\ref{proof_miso}. 
For $\bar{B} > 0$, the rate loss due to finite
feedback is a constant.
As $\bar{B} \to 0$, this rate loss tends to infinity, since
with $\bar{B}=0$, the capacity tends to a constant as
$N_t \to \infty$, whereas the capacity grows as $\log N_t$ 
for $\bar{B} > 0$.
Of course, as $\bar{B} \to \infty$ (unlimited feedback), 
the rate loss vanishes.

RVQ is asymptotically optimal in the following sense.  
Suppose that $\{ \mathcal{V}_{N_t} \}$ is an arbitrary 
sequence of codebooks for the beamforming vector where 
\begin{equation}
  \mathcal{V}_{N_t} = \left\{ \bm{v}_1^{N_t},
  \bm{v}_2^{N_t}, \ldots, \bm{v}_{2^B}^{N_t} \right\}
\end{equation}
is the codebook for a particular $N_t$
and $\|\bm{v}_j^{N_t} \|^2 = 1$ for each $j$. 
The associated rate is given by
\begin{equation}
  I_{\mathcal{V}_{N_t}} = \max_{1 \le j \le 2^B} 
       \log (1 + \rho | \bm{h}^{\dag} \bm{v}_j^{N_t} |^2 )
\end{equation}
and the rate difference $I^{\triangle}_{\mathcal{V}_{N_t}} =
I_{\mathcal{V}_{N_t}} - \log(\rho N_t)$.
\begin{theorem}
\label{beam_opt}
 For any sequence of codebooks $\{ \mathcal{V}_{N_t} \}$, 
\begin{equation} 
\limsup_{(N_t,B) \to \infty} 
E_{\bm{h}} [ I^{\triangle}_{\mathcal{V}_{N_t}}] \le
 \mathcal{I}_{\rvq}^{\triangle} .
\end{equation}
\end{theorem}	
The proof is given in Appendix \ref{proof_opt_miso}.

Although the optimality of RVQ holds only in the large system limit,
numerical results in Section \ref{num_results} show that for
finite-size systems of interest RVQ performs essentially the same as
optimized quantization codebooks.

%----------------------------------------------------
\subsection{Multi-Input Multi-Output (MIMO) Channel}
%----------------------------------------------------

We now consider quantized beamforming for a MIMO channel, i.e., 
with multiple transmit {\em and} receive antennas.
Taking the rank $K=1$ maximizes the diversity gain \cite{ZhengTse},
but the corresponding capacity grows only as $\log N_t$
instead of linearly with $N_t$, which is the case
when $K$ grows proportionally with $N_t$.
(This is true with both
unlimited and limited feedback, assuming a fixed number
of feedback bits per precoder element.)  
Also, a beamformer is significantly less complex 
than a matrix precoder with $K>1$, and requires
less feedback to specify.

We again consider an RVQ codebook $\mathcal{V}$ with $2^B$
independent unit-norm vectors, where each vector is uniformly
distributed over the $N_t$-dimensional unit sphere.  
The achievable rate is $E_{\bH} [ I_{\rvq}^{N_t} ]$, where
\begin{eqnarray}
I_{\rvq}^{N_t} 
  & = & E_{\mathcal{V}} \left[
        \left. \max_{1 \le j \le 2^B} \log \left( 1 + \rho 
	\| \bm{H} \bm{v}_j \|^2
	\right) \right| \bm{H}\right] \\
  & = & E_{\mathcal{V}} \left[ \log ( 1 + \rho \left.
       \max_{1 \le j \le 2^B} \| \bm{H} \bm{v}_j \|^2 )\right| \bm{H}
       \right] .
\end{eqnarray}
As for the MISO channel, with unlimited feedback the 
achievable rate increases as $\log (\rho N_t)$.
We again define the rate difference due to quantization as
\begin{equation}
  I^{\triangle}_{\rvq} \triangleq I_{\rvq}^{N_t} - \log(\rho N_t) =
  E_{\mathcal{V}} \left[ \left. \log \left( \frac{1}{\rho N_t} +
  \max_j \gamma_j
  \right) \right| \bm{H} \right]
\label{IDr}
\end{equation}
where
\begin{equation}
\gamma_j = \frac{1}{N_t} \bm{v}_j^\dag \bm{H}^\dag \bm{H} \bm{v}_j .
\label{eq:gammaj}
\end{equation}
Evaluating the expectation in \eqref{IDr} is difficult for finite
$N_t$, $N_r$, and $B$, so that we again resort to a large system
analysis. Namely, we let $N_t$, $N_r$, and $B$ each tend to infinity
with fixed $\B = B/N_t$ and $\N = N_r/N_t$.  For each $N_t$ and $N_r$
the channel matrix $\bm{H}$ is chosen as the $N_r \times N_t$
upper-left corner of a matrix $\bar{\bm{H}}$ with an infinite number
of rows and columns, and with {\em i.i.d.} complex Gaussian entries.

The received power in this large system limit is given by
\begin{equation}
  \gamma^{\infty}_{\rvq} = \lim_{(N_t, N_r, B) \to \infty}
    \left[ \left. \max_{1 \le j \le 2^B} \gamma_j
    \right| \bar{\bm{H}} \right]
\label{vrr}
\end{equation}
where convergence to the deterministic limit
can be shown in the mean square sense.
Conditioned on $\bar{\bm{H}}$, the $\gamma_j$'s are {\em
i.i.d.} since the beamforming vectors $\bm{v}_j$ are {\em i.i.d.},
and applying \cite[Theorem 2.1.2]{galambos}, it
can be shown that
\begin{equation}
  \gamma^{\infty}_{\rvq} = \lim_{(N_t, N_r, B) \to \infty} F^{-1}_{\gamma |
  \bar{\bm{H}}} \left( 1 - 2^{-B} \right)
\label{gir}
\end{equation}
where $F_{\gamma | \bar{\bm{H}}} ( \cdot )$ is the cdf of $\gamma_j$
given $\bar{\bm{H}}$.  Analogous results for the interference power in
CDMA with quantized signatures have been presented in \cite{cdma04},
so that we omit the proofs of \eqref{vrr} and \eqref{gir}.  Note that
$\N \leq \gamma^{\infty}_{\rvq} \leq (1 + \sqrt{\N})^2$, where the
lower and upper bounds correspond to $\B = 0$ and $\B = \infty$,
respectively.  That is, $ (1 + \sqrt{\N})^2$ is the asymptotic maximum
eigenvalue of the channel covariance matrix $\frac{1}{N_t}
\bm{H}^{\dag}\bm{H}$ \cite{marcenko}.  The asymptotic rate difference
is given by
\begin{equation}
  \mathcal{I}_{\rvq}^{\Delta}  =  \lim_{(N_t, N_r, B) \to \infty}
  I^{\Delta}_{\rvq} =  \log ( \gamma^{\infty}_{\rvq} )
\end{equation}

The limit in \eqref{gir} can be explicitly evaluated, 
and is independent of the channel realization $\bar{\bm{H}}$.

\begin{theorem}
\label{ua}
For $0 \le \B \le \B^*$, $\srvq$ satisfies
\begin{equation}
  \left( \srvq \right)^{\N} \me^{-\srvq} = 2^{-\B} \left(
  \frac{\N}{\me} \right)^{\N}
\label{lsr}
\end{equation}
 and for $\B \ge \B^*$,
\begin{equation}
\begin{split}
  \srvq &= (1 + \sqrt{\N})^2 - \exp \big\{ \frac{1}{2} \N \log(\N) \\
   &\quad - (\N-1) \log (1 + \sqrt{\N}) + \sqrt{\N} - \B \log(2) \big\}
\end{split}
\label{g1N}
\end{equation}
where 
\begin{equation}
\B^* = \frac{1}{\log(2)}
\left( \N\log \left(\frac{\sqrt{\N}}{1+\sqrt{\N}} \right) + \sqrt{\N} \right).
\label{bst}
\end{equation}
\end{theorem}

The proof is given in Appendix~\ref{proof_mimo} and is motivated by an
analogous result for CDMA, presented in \cite{dai_ciss}.  As stated in
Theorem \ref{ua}, $\srvq$ depends only on $\B$ and $\N$.  Letting $\N
\to 0$ gives the the asymptotic capacity of the MISO channel with RVQ.
As for the MISO channel, RVQ is asymptotically optimal.

\begin{theorem}
\label{mnb}
 As $(N_t,N_r,B) \to \infty$ with fixed $\N = N_r/N_t$ and $\B = B/N_t$,
 \begin{equation}
   \limsup_{(N_t,N_r,B) \to \infty} I_{\mathcal{V}_{N_t}}^{N_t} -
   \log(\rho N_t) \le \mathcal{I}_ {\rvq}^{\Delta}
 \end{equation}
for any sequence of codebooks $\{ \mathcal{V}_{N_t} \}$.
\end{theorem}
The proof is similar to the proof of Theorem~2 in \cite{cdma04} 
and is therefore omitted.

%-----------------------------
\subsection{Numerical Results}
%-----------------------------
\label{num_results}

\begin{figure}
\centering
\includegraphics[width=3.25in]{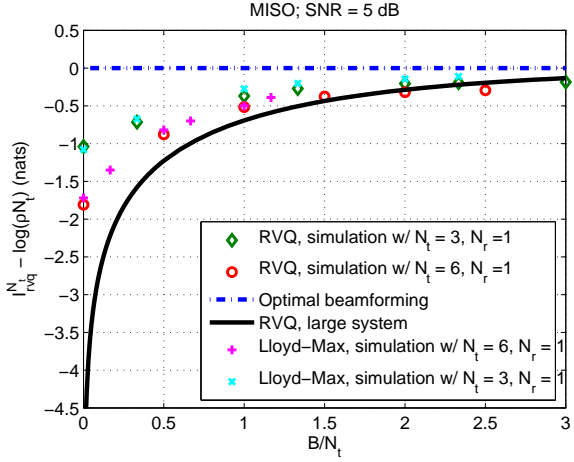}
\caption{Asymptotic and simulated rate differences versus feedback bits
  for a MISO channel with beamforming.}
\label{beam_miso}
\end{figure}

\begin{figure}
\centering
\includegraphics[width=3.25in]{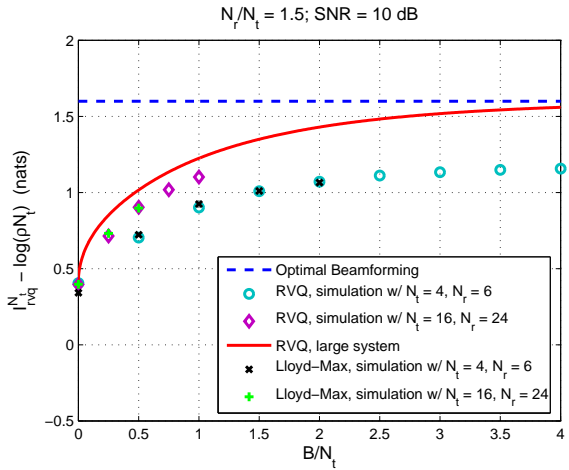}
\caption{Asymptotic and simulated rate differences versus feedback bits
  for a MIMO channel with beamforming ($\N = 1.5$).}
\label{beam_mimo}
\end{figure}

Figs.~\ref{beam_miso} and~\ref{beam_mimo} show
$\mathcal{I}_{\rvq}^{\triangle}$ for MISO and MIMO channels,
respectively, with beamforming and RVQ versus normalized feedback bits
($\bar{B}$) with $\rho = 5$ and 10 dB.  Also shown for comparison are
achievable rates with a quantization codebook optimized via the
Lloyd-Max algorithm \cite{narula98,lau04,roh_it06}, and the capacity
with perfect beamforming, corresponding to unlimited feedback. The
results for RVQ are averaged over codebook realizations, and are
essentially the same as those shown for the optimized Lloyd-Max
codebooks.  For the MISO channel, the asymptotic capacity
\eqref{eq:miso} accurately predicts the simulated results shown even
with a relatively small number of transmit antennas ($N_t = 3 \text{
and } 6 $).  For the MIMO results $\N = 1.5$, and simulation results
are shown for $4 \times 6$ and $16 \times 24$ channels.  The
asymptotic results accurately predict the performance for the larger
channel, and are somewhat less accurate for the smaller channel.

Comparing finite feedback with perfect beamforming, the results show
that one feedback bit per complex entry ($\B = 1$) provides more than
50\% of the potential gain due to feedback.  For both the MISO and
MIMO examples shown, the perfect beamforming capacity is nearly
achieved with two feedback bits per complex coefficient.

%%%%%%%%%%%%%%%%%%%%%%%%%%%%%%%%%%%%%%%%%%%%%%%
\section{Precoding Matrix with Arbitrary Rank}
\label{opt}
%%%%%%%%%%%%%%%%%%%%%%%%%%%%%%%%%%%%%%%%%%%%%%%

In this section we consider the performance of a single-user MIMO
channel with precoding matrix $\bV$ having rank $K > 1$.  We wish
to determine the asymptotic capacity with RVQ as in the
previous section.  Here we consider the large system limit
as $ ( N_t, N_r, B, K ) \to \infty $ with fixed ratios 
$ \N = N_r/N_t $, $ \hat{B} = B/N_r^2 $, and $ \bar{K} = K/N_t$. 
That is, we scale the rank of the precoding matrix with $N_t$.
The number of feedback bits is normalized by $N_r^2$, 
instead of $N_r$, since the feedback must
scale linearly with degrees of freedom (in this case the number of
channel elements $N_r N_t$).)  Given a fixed number of
feedback bits per channel coefficient, 
the capacity grows {\em linearly} with
the number of antennas ($N_t$ or $N_r$).

Given a rank $ K \le N_t $, the precoding matrix 
is chosen from the RVQ set
\begin{equation}
  \mathcal{V} = \{ \bV_j, 1 \le j \le 2^B \},
\end{equation}
where the entries are independent $ N_t \times K $ random unitary
matrices, i.e., $ \bV_j^{\dag} \bV_j = \bm{I}_K $.  
This codebook is an extension of the RVQ codebook for beamforming.  
Letting
\begin{equation}
J^{N_r}_j = \frac{1}{N_r} \log \det 
\left( \bm{I}_{N_r} + \frac{\rho}{K} \bm{H}
\bV_j \bV_j^{\dag} \bm{H}^{\dag} \right) ,
\end{equation}
the receiver again selects the quantized precoding matrix, 
which maximizes the mutual information
\begin{equation}
  \hat{\bV} = \arg \max_{\quad 1 \le j \le 2^B} J^{N_r}_j .
\label{hatv}
\end{equation}
For finite $N_r$, we define
\begin{align}
I_{\rvq}^{N_r} & = E_{\mathcal{V}} 
[ \max_{\quad 1 \le j \le 2^B} J^{N_r}_j | \bar{\bH} ]\\
&= E_{\mathcal{V}} \left[ \left. \frac{1}{N_r} \log \det
       \left( \bm{I}_{N_r} + \frac{\rho}{K} \bm{H} \hat{\bV}
       \hat{\bV}^{\dag} \bm{H}^{\dag} \right) \right| \bar{\bH} \right]
\label{eq:i_n}
\end{align}
and the average sum mutual information per receive antenna 
with $ B $ feedback bits is then $E_{\bH} [ I_{\rvq}^{N_r} ] $.

Here the power allocation over channel modes is ``on-off''.
Namely, active modes are assigned equal powers.
This simplifies the analysis, and it has been observed
that the additional gain due to an optimal power allocation
(water pouring) is quite small \cite{dai_liu}.

Since the entries of the RVQ codebook are {\em i.i.d.}, the mutual
informations $J^{N_r}_j, j= 1, \ldots, 2^B$, are also {\em i.i.d.} for
a given $\bm{H}$.  In principle, the large system limit of
$I_{\rvq}^{N_r}$ can be evaluated, in analogy with \eqref{gir}, given
the cdf of $J^{N_r}_j $ given $\bm{H}$, denoted as $F_{J;N_r |
\bm{H}}$.  This cdf appears to be difficult to determine in
closed-form for general $(N_r,N_t,K)$, so that we are unable to derive
the exact asymptotic capacity with RVQ. Still, we can provide an
accurate approximation for this large system limit.  Before presenting
this approximation, we first compare the capacity with no channel
information at the transmitter ($ \hat{B} = 0 $) to the capacity with
perfect channel information ($ \hat{B}= \infty $).

If $\hat{B}=0$, then the optimal transmit covariance matrix 
$ \bV \bV^{\dag} = \bm{I}_{N_t} $ and $K = N_t$ \cite{telatar99}.  
That is, all channel modes are
allocated equal power.  As $ (N_t, N_r) \to \infty $
with fixed $ \N = N_r/N_t $, the capacity per 
receive antenna is given by
\begin{align}
   \frac{1}{N_r} \log \det \left( \bm{I}_{N_r} + \frac{\rho}{N_t} \bm{H}
   \bm{H}^{\dag} \right) 
   & \to \int_0^{\infty} \log \left( 1 + \rho \lambda \right) 
     g(\lambda) \, \diff \lambda \label{int} \\ 
   & = \mathcal{I}_{\rvq} ( \hat{B} = 0)
\end{align}
where convergence is in the almost sure sense, 
and $ g(\lambda) $ is the asymptotic probability density function 
for a randomly chosen eigenvalue of 
$ \frac{1}{N_t} \bm{H} \bm{H}^{\dag} $, and is given by
\cite{marcenko}
\begin{gather}
\label{eq:marcenko-pastur1}
  g(\lambda)  =  \frac{ \sqrt{(\lambda - a)(b - \lambda)} }{ 2 \pi
      \lambda \N } \quad {\mathrm{for}} \quad a \le \lambda \le b, \\
  a =  \left( 1 - \sqrt{\N} \right)^2 \quad 
     {\mathrm{and}} \quad
  b =  \left( 1 + \sqrt{\N} \right)^2
\label{eq:marcenko-pastur2}
\end{gather}
for $\N \le 1$.
The integral in \eqref{int} has been evaluated in \cite{rapajic00},
which gives the closed-form expression
\begin{equation}
  \mathcal{I}_{\rvq} ( \hat{B} = 0) = \log \rho y + \frac{1 -
    \N}{\N} \log \left( \frac{1}{1 - z} \right) -
    \frac{z}{\N}
\end{equation}
where
\begin{align}
   y & =  \frac{1}{2} \left( 1 + \N + \frac{1}{\rho} + \sqrt{
         \left( 1 + \N + \frac{1}{\rho} \right)^2 - 4\N } 
         \right) \\
   z & =  \frac{1}{2} \left( 1 + \N + \frac{1}{\rho} - \sqrt{
         \left( 1 + \N + \frac{1}{\rho} \right)^2 - 4\N } \right).
\end{align}

If $ \hat{B} = \infty $, then the $K$ columns of the optimal $ \bV$
are the eigenvectors of the channel covariance matrix corresponding to
the $K$ largest eigenvalues.  As $( N_t, N_r, B ) \to \infty $, we
have
\begin{equation}
   \mathcal{I}_{\rvq} (\hat{B} = \infty) = \int_\eta^{\infty}
   \log \left( 1 + \frac{\rho}{\K} \lambda \right) g(\lambda) 
   \,\diff \lambda
\label{wt2}
\end{equation}
where $ \eta $ satisfies
\begin{equation}
   \int_{\eta}^\infty g(\lambda) \,\diff \lambda = \min\{1, \frac{\K}{\N}\}
\label{wt1}
\end{equation}
for $\N \le 1$. We emphasize that this corresponds to a uniform
allocation of power over the set of $K$ active eigenvectors.  (This
result has also been presented in \cite{dai_liu}.) The rank of the
optimal $\bV$, or optimal multiplexing gain, is at most $ \min \{ N_t,
N_r \} $ and can be obtained by differentiating \eqref{wt2} with
respect to $\K$. It can be verified that $ \mathcal{I}_{\rvq}(\hat{B}
= 0) \le \mathcal{I}_{\rvq}(\hat{B} = \infty) $.

\begin{figure}[ht]
\centering 
\includegraphics[width=3.25in]{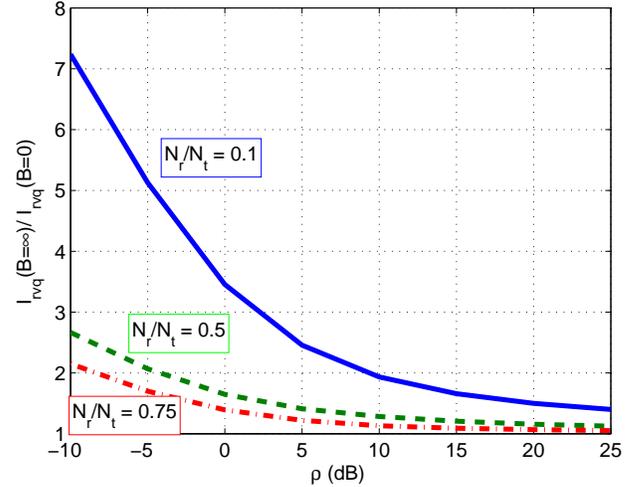}
\caption{The rate ratio $\mathcal{I}_{\rvq}(\hat{B} = \infty) /
\mathcal{I}_{\rvq}(\hat{B} = 0)$ versus SNR (dB) for various values of
$\N$.}
\label{cap_ratio}
\end{figure}

To illustrate the increase in capacity with feedback,
in Fig.~\ref{cap_ratio} we plot the rate ratio 
$\mathcal{I}_{\rvq}(\hat{B} = \infty)/
\mathcal{I}_{\rvq}(\hat{B} = 0)$
versus SNR for different values of $\N$,
where $\mathcal{I}_{\rvq}(\hat{B} = \infty )$ 
is optimized over rank $K$.
For large SNR $\rho$, we can expand
\begin{align}
  \mathcal{I}_{\rvq}(\hat{B} = 0) & = \log(\rho) + o(\log(\rho))\\
  \mathcal{I}_{\rvq}(\hat{B} = \infty) & = \log(\rho)
  \int_{\eta}^{\infty} g(\lambda) \, \diff \lambda + o(\log(\rho)) .
\end{align}
Therefore
\begin{equation}
   \lim_{\rho \to \infty} \frac{\mathcal{I}_{\rvq}(\hat{B} =
    \infty)}{\mathcal{I}_{\rvq}(\hat{B} = 0)} 
    = \int_{\eta}^{\infty} g(\lambda)\, \diff \lambda \\
    = \min\{1, \frac{\K}{\N}\}
\end{equation}
which implies that the optimal rank $K^* = \min \{ N_t, N_r \}$, 
and the corresponding asymptotic rate ratio is one. 
The increase in
achievable rate from feedback is small in this case,
since for large SNRs, the transmitter excites all channel
modes, and the uniform power allocation asymptotically gives the same
capacity as water pouring.  Of course, although the increase in rate
is small, feedback can simplify coding and decoding. 

For small $\rho$, we can expand $\log(1 + \rho \lambda)$ 
and $\log(1 + \rho \lambda /\K)$ in Taylor series. 
Taking $\rho \to 0$ gives
\begin{align}
   \lim_{\rho \to 0} & \frac{\mathcal{I}_{\rvq}(\hat{B} =
   \infty)}{\mathcal{I}_{\rvq}(\hat{B} = 0)}  \nonumber\\
& =  \frac{1}{\K}
   \int_{\eta}^{\infty} \lambda g (\lambda) \, \diff \lambda
   \le \frac{1}{\N}
   \frac{\int_{\eta}^\infty \lambda g(\lambda) \,\diff
   \lambda}{\int_{\eta}^\infty g(\lambda) \,\diff \lambda}
\label{eq:rateratio2} \\
& \le  \frac{1}{\N}
   \frac{b \int_{\eta}^b g(\lambda) \,\diff
   \lambda}{\int_{\eta}^b g(\lambda) \,\diff \lambda}
   = \left( 1 + \frac{1}{\sqrt{\N}} \right)^2
\label{eq:rateratio3}
\end{align}
where $b$ is the asymptotic maximum eigenvalue given by
\eqref{eq:marcenko-pastur2}.  The inequality in \eqref{eq:rateratio2}
follows from \eqref{wt1}, which implies $ \K \ge \N \int_{\eta}^\infty
g(\lambda) \,\diff \lambda $. Note that (\ref{eq:rateratio3})
corresponds to allocating all transmission
power to the strongest channel mode, which is known
to maximize capacity at low SNRs.
The maximal rate ratio~\eqref{eq:rateratio3} can also be 
obtained from Theorem~\ref{ua}.

The rate increase due to feedback is substantial when $\N$ is small,
and the rate ratio tends to infinity as $\N \to 0$.
This is because the channel becomes a MISO channel, in which case
the capacity is a constant with $\B=0$ and increases
as $\log (\rho N_t)$ with $\B = \infty$.

To evaluate the asymptotic capacity with arbitrary $\hat{B}$,
we approximate $J_j^{N_r}$ given $\bar{\bm{H}}$ 
as a Gaussian random variable. This is motivated
by the fact that $J_j^{N_r}$ is Gaussian in the large
system limit \cite{bai04}, since $\bH \bV$ is {\em i.i.d.}
Conditioning on $\bm{H}$ introduces dependence
among the elements of $\bH \bV$; however, numerical examples
indicate that the Gaussian assumption is still valid for large $N_t$
and $N_r$. Alternatively, if we do not condition 
on $\bH$, then the rates $\{J_j^{N_r}\}$ are dependent.
Application of the results
from extreme statistics, assuming
the rates $\{J_j^{N_r}\}$ are independent, gives
an upper bound on the asymptotic achievable rate
(e.g., see the proof of Theorem 2 in \cite{cdma04}).
This is illustrated by subsequent numerical results.

Evaluating the large system limit of $I_{\rvq}^{N_r}$,
assuming that the cdf of $J_j^{N_r}$ is Gaussian, gives
the approximate rate
\begin{equation}
  \tilde{\mathcal{I}}_{\rvq} = \mu_J + \sigma_J \sqrt{ 2
  \hat{B} \log 2 }
\label{appxm}
\end{equation}
independent of the channel realization, where $\mu_J$ and $\sigma_J^2$
are the asymptotic mean of $J_j^{N_r}$, and variance of
$N_r^2 J_j^{N_r}$, respectively.  The derivation
of \eqref{appxm} is a straightforward extension of
\cite[Sec. 2.3.2]{galambos} and is not shown here.  As $ \hat{B} \to 0
$, this approximation becomes exact.  However, as $ \hat{B} \to \infty
$, the approximate rate $ \tilde{\mathcal{I}}_{\rvq} \to \infty $,
whereas the actual rate $ \mathcal{I}_{\rvq} (\hat{B} = \infty) $ is
finite, and can be computed from \eqref{wt2} and \eqref{wt1}.  This is
because $ J_j^{N_r}$ is bounded for all $N_r$, whereas a Gaussian
random variable can assume arbitrarily large values.  Therefore the
Gaussian approximation gives an inaccurate estimate of $
\mathcal{I}_{\rvq} $ for large $ \hat{B} $.  (This implies that we
should approximate $\mathcal{I}_{\rvq}$ as $\min \{
\tilde{\mathcal{I}}_{\rvq}, \mathcal{I}_{\rvq} (\hat{B} =\infty) \}$.)

The asymptotic mean and variance of $J_j^{N_r}$ are computed in
Appendix~\ref{lemma_gauss}. The asymptotic mean is given by
\begin{equation}
\label{mu}
\begin{split}
   \mu_J = \frac{\K}{\N} \log \left( 1 + \frac{\N}{\K} \rho -
   \frac{\N}{\K} \rho v \right) \\
    + \log \left( 1 + \rho - \frac{\N}{\K}
   \rho v \right) - v
\end{split}
\end{equation}
where
\begin{equation}
   v = \frac{1}{2} + \frac{\K}{2\N} + \frac{\K}{2 \N \rho} -
      \frac{1}{2} \sqrt{ \left(1 + \frac{\K}{\N} + \frac{\K}{\N \rho}
      \right)^2 - \frac{4 \K}{\N}} .
\end{equation}
The asymptotic variance is approximated for $0 \le \K = \N \le
1$ and small SNR ($\rho \le -5\, \text{dB}$) as
\begin{equation} 
\label{ll1}
   \sigma_J^2 \approx \rho^2 (1 - \N) .
\end{equation}
The asymptotic variance for moderate SNRs and normalized rank
$\K \ne \N$ can be computed easily via 
numerical simulation.\footnote{We note that the simulation
needed to compute this variance is much simpler than the simulation, 
which would be required to obtain the RVQ rate directly, especially
with a moderate to large number of feedback bits.}

In contrast with the beamforming results in the preceding section, we
are unable to show that RVQ is asymptotically optimal when the
precoding matrix has arbitrary rank.  The corresponding argument for
beamforming relies on the evaluation of the asymptotic rate difference
$\mathcal{I}_{\rvq}^\Delta$. Since here we are unable to evaluate
$\mathcal{I}_{\rvq}$ exactly, we cannot apply that
argument. Nevertheless, numerical results have indicated that the
performance of RVQ matches that of optimized codebooks (e.g., see
\cite{commag04}).  

\begin{figure}[ht]
\centering 
\includegraphics[width=3.25in]{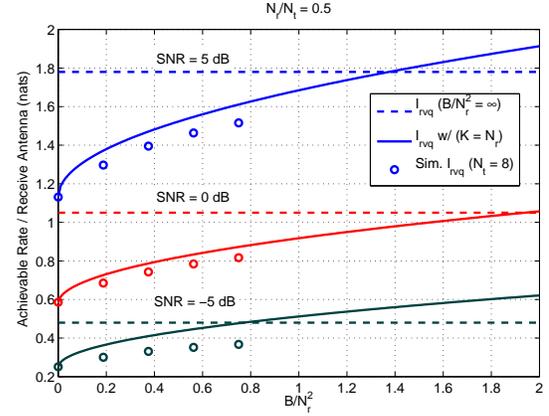}
\caption{Sum mutual information per receive antenna with RVQ and an
  optimal receiver versus normalized feedback.  The asymptotic
  approximation is shown along with Monte Carlo simulation results for
  an $8 \times 4$ channel.}
\label{rbo}
\end{figure}

Fig.~\ref{rbo} shows $ \tilde{\mathcal{I}}_{\rvq} $ with normalized
rank $ \K = \N $ versus $ \hat{B} $ for $ \rho = -5, 0, 5 \ \text{dB}$
and $ \N = 0.5 $.  The dashed lines show the unlimited feedback
capacity $ \mathcal{I}_{\rvq} (\hat{B} = \infty) $, which is computed
from \eqref{wt2} with optimized $\K$.  
The asymptotic rate with RVQ is computed from
\eqref{appxm}, where $\sigma_J$ for $\rho = -5 \, \text{dB}$ is
approximated by \eqref{ll1}, and $\sigma_J$ is determined
from simulation with $N_t = 20$ 
for $\rho = 0 \text{ and } 5 \text{ dB}$.
Also shown in Fig.~\ref{rbo} are simulation results for 
$ \mathcal{I}_{\rvq} $ with $N_t = 8$ and $N_r = 4$. 
Because the size of the RVQ codebook increases exponentially
with $\hat{B}$, it is difficult to generate simulation
results for moderate to large values of $\hat{B}$.
Hence simulation results are shown only for $\hat{B} \leq 0.8$.
The asymptotic results accurately approximate the 
simulated results shown. The accuracy increases
as the feedback $\hat{B}$ decreases.

\begin{figure}
\centering 
\includegraphics[width=3.25in]{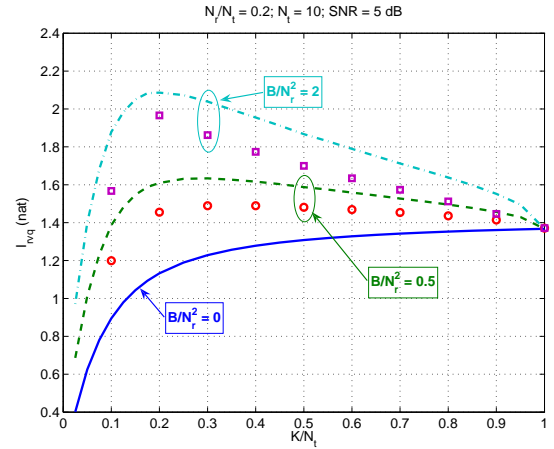}
\caption{Mutual information per receive antenna versus normalized rank
  with different normalized feedback. Discrete points correspond
  to simulation with $N_t = 10$.}
\label{rro}
\end{figure}

Since $ \tilde{\mathcal{I}}_{\rvq} $ is a function of both rank $ \K $
and feedback $ \hat{B} $, for a given $\hat{B}$, we can select $ \K$
to maximize $ \tilde{\mathcal{I}}_{\rvq} $.  Fig.~\ref{rro} shows
mutual information per receive antenna versus normalized rank from
\eqref{appxm} with $ \N = 0.2 $, $ \rho = 5 $ dB, and different values
of $ \hat{B} $. ($\sigma_J$ is obtained from numerical simulations.)
The maximal rates are attained at $ \K = 1$, $0.3$, and $0.2$ for
$\hat{B} = 0$, $0.5$, and $2$, respectively.  In general, the optimal
rank is approximately $\N$ for large enough $\hat{B}$ and SNR.  The
results in Fig.~\ref{rro} indicate that taking $\K=\N$ achieves
near-optimal performance, independent of $\hat{B}$ when $\hat{B}>0$.
As $ \hat{B} $ increases, the rate increases and the difference
between the rate with optimized rank and full-rank ($\K=1$) also
increases.  For the example shown, the rate increase from selecting
the optimal rank is as high as $50\%$ when $\hat{B} = 2$.

%%%%%%%%%%%%%%%%%%%%%%%%%%%%%%%%%%%%%%%%%%%%%%%%%%%%%
\section{Quantized Precoding with Linear Receivers}
\label{linear}
%%%%%%%%%%%%%%%%%%%%%%%%%%%%%%%%%%%%%%%%%%%%%%%%%%%%%

In this section we evaluate the performance
of a quantized precoding matrix with linear receivers
(matched filter and MMSE), and compare with
the performance of the optimal receiver. 
As $\hat{B} \to \infty$, the optimal precoding matrix
eliminates the cross-coupling among channel modes, 
and the optimal receiver becomes the linear matched filter. 
Hence the corresponding achievable rates should be the same
in this limit. However, for finite $\hat{B}$ the optimal receiver
is expected to perform better than the linear receiver.
Given a target rate, increasing the feedback
therefore enables a reduction in receiver complexity.

We again assume that there are $K$
independent data streams, which are multiplexed by the linear precoder
onto $N_t$ transmit antennas.  To detect the transmitted
symbols in data stream $k$, the received signal $\bm{y}$ is passed
through the $N_r \times 1$ receive filter $\bm{c}_k$.
The matched filter is given by
\begin{equation}
\label{mfr}
  \bm{c}_k = \frac{1}{\sqrt{K}} \bm{H} \bm{v}_{k}
\end{equation}
where $\bm{v}_{k}$ is the $k$th column of the precoding matrix
$\bV$, and the linear MMSE filter is given by
\begin{equation}
\label{mmser}
  \bm{c}_k = \frac{1}{\sqrt{K}} \left( \frac{1}{K} \bm{H} \bV
  \bV^{\dag} \bm{H}^{\dag} + \varn \bm{I}_{N_r} \right)^{-1} \bm{H}
  \bm{v}_{k} .
\end{equation}
The SINR at the output of the linear filter $\bm{c}_k$ is
\begin{equation}
\label{sinr}
  \mathrm{{\sf SINR}}_k = \frac{| \bm{c}_k^{\dag} \bm{H} \bm{v}_{k}
  |^2}{\bm{c}_k^{\dag} \left( \sum_{i \ne k} \bm{H} \bm{v}_{i}
  \bm{v}_{i}^{\dag} \bm{H}^{\dag} + K \varn \bm{I}_{N_r} \right) \bm{c}_k} .
\end{equation}
Of course, the interference among data streams can significantly 
decrease the channel capacity.

The performance measure is again mutual information between the
transmitted symbol $x_k$ and the output of the filter $\bm{c}_k$,
denoted by $\hat{x}_k$.  
In what follows, we assume independent coders and decoders for
each data stream.
Assuming that the interference plus noise at
the output of the linear filter has a Gaussian distribution,
which is true in the large system limit to be considered, 
the sum mutual information of all data streams 
per receive antenna is given by
\begin{eqnarray}
  R^{N_r} & = & \frac{1}{N_r} \sum_{k = 1}^K I( x_k, \hat{x}_k) \\ 
          & = & \frac{1}{N_r} \sum_{k = 1}^K 
                \log (1 + \gamma_k) .
\label{sumr}
\end{eqnarray}
where $\gamma_k$ is the SINR for the $k$th data stream.
Given a channel matrix $\bm{H}$, the sum rate $R^{N_r}$
depends on the precoding matrix $\bV$.  We are interested in
maximizing $R^{N_r}$ subject to the power constraint
$ \| \bm{v}_{k} \| \le 1\mathrm{,} \ \forall k$, assuming that the
power is allocated equally across streams.

Given the codebook of precoding matrices
$\mathcal{V} = \{ \bV_j, \ 1 \le j \le 2^B \}$,
the receiver selects the precoding matrix 
\begin{equation}
  \hat{\bV} = \arg \max_{1 \le j \le 2^B} R^{N_r}( \bV_j ) .
\label{vq}
\end{equation}
We again consider RVQ, in which
the $\bV_j$'s are {\em i.i.d.} unitary matrices.

%--------------------------
\subsection{Matched filter}
%--------------------------

Substituting \eqref{mfr} into \eqref{sinr}, the SINR at the output
of the matched filter is given by
\begin{equation}
  \gamma_{k;\mf} = \frac{(\bm{v}_{k}^{\dag} \bm{H}^{\dag} \bm{H}
  \bm{v}_{k})^2}{K \varn ( \bm{v}_{k}^{\dag} \bm{H}^{\dag}
  \bm{H} \bm{v}_{k}) + \sum_{i = 1, i \ne k}^K |
  \bm{v}_{k}^{\dag} \bm{H}^{\dag} \bm{H} \bm{v}_{i} |^2}
\end{equation}
where subscript $k$ denotes the $k$th data stream.  
The average sum rate per receive antenna is given by
\begin{equation}
  E_{\bH, \mathcal{V}} \left[ \max_{1 \le j \le 2^B} \{ R_{\mf}^{N_r} (\bV_j)
  = \frac{1}{N_r} \sum_{k = 1}^K \log( 1 + \gamma_{k;\mf}) \} \right]
\label{eq:mf}
\end{equation}
where the expectation is over the channel matrix and codebook.  
Since the pdf of $R_{\mf}^{N_r}$ is unknown
for finite $(N_t, N_r, K)$, we are unable
to evaluate \eqref{eq:mf}.
Motivated by the central limit theorem,\footnote{The 
terms in the sum in \eqref{eq:mf} are not {\em i.i.d.}, which
prevents a direct application of the central limit theorem.}
in what follows we approximate the cdf of 
$R_{\mf}^{N_r}$ as Gaussian. 
The mean is taken to be the asymptotic limit
\begin{equation}
  \mu_{\mf} = \lim_{(N_t,N_r,K) \to \infty} R_{\mf}^{N_r} =
  \frac{\K}{\N} \log \left( 1 + \frac{\N}{\K (1 + \varn)} \right) .
\label{mumf}
\end{equation} 
This limit follows from the fact that $\gamma_{k;\mf}$ converges
almost surely to $[ \K(1 + \varn)/\N ]^{-1}$ as $(N_t,N_r,K) \to
\infty$ with fixed $\N$ and $\K$. As for the optimal receiver,
\begin{equation}
  N_r^2 \var [ R_{\mf}^{N_r} | \bV_j ] \to \sigma^2_{\mf}
\label{sm2f}
\end{equation}
where $\sigma^2_{mf}$ can be easily obtained by numerical simulations.

\begin{figure}
  \centering
  \includegraphics[width=3.25in]{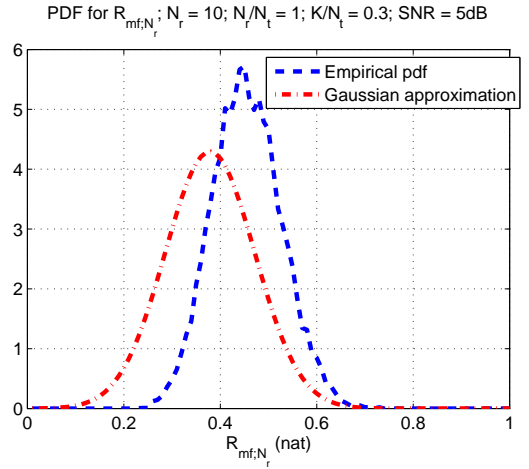}
  \caption{Comparison of the empirical pdf for $R_{\mf;N_r}$ with the
  Gaussian approximation.}
  \label{mf_pdf}
\end{figure}

The accuracy of the Gaussian approximation for $R_{\mf}^{N_r}$ is
illustrated in Fig.~\ref{mf_pdf}, which compares the empirical pdf
with the Gaussian approximation for $N_r = 10$, $\N = 1$, $K/N_r =
0.3$ and $\mathrm{SNR} = 5 \, \mathrm{dB}$.  The difference between
the empirical and asymptotic means vanishes as $(N_t,N_r,K) \to
\infty$.

We wish to apply the theory of extreme order
statistics \cite{galambos} to evaluate the large system limit
\begin{equation}
\mathcal{R}_{\rvq;\mf} = 
\lim_{(N_t,N_r,K,B) \to \infty}
[ \max_{1 \le j \le 2^B} R_{\mf}^{N_r} (\bV_j) | \mathcal{V}].
\end{equation}
Given $\mathcal{V}$, the sum rates $\{ R_{\mf}^{N_r} (\bV_1), \ldots,
R_{\mf}^{N_r} (\bV_{2^B}) \}$ are identically distributed.  However,
the $R_{\mf}^{N_r} (\bV_j)$'s are not independent since each depends
on $\bH$.  This makes an exact calculation of the asymptotic rate
difficult.  Nevertheless, for a small number of entries in the
codebook (small $B$), assuming that the rates for a given codebook are
independent leads to an accurate approximation. We therefore replace
the rates $R_{\mf}^{N_r} (\bV_j)$, $j=1,\cdots,2^B$, with {\em i.i.d.}
Gaussian variables with mean $\mu_{\mf}$ and variance $\sigma^2_{\mf}
/ N_r^2$.  In analogy with the analysis of the optimal receiver in the
preceding section, this gives the approximate asymptotic rate
\begin{equation}
          \tilde{\mathcal{R}}_{\rvq;\mf}
          = \mu_{\mf} + \sigma_{\mf} \sqrt{2 \hat{B} \log 2} .
\label{rvqmf}
\end{equation}
Numerical results, to be presented, show that this asymptotic
approximation is very accurate for small to moderate values of
normalized feedback $\hat{B}$.  As $\hat{B} \to 0$, this approximation
becomes exact.  However, as $ \hat{B} \to \infty $, $
\tilde{\mathcal{R}}_{\rvq;\mf} \to \infty $, whereas $
\mathcal{R}_{\rvq;\mf}$ with $\hat{B} = \infty$ is the same as the
asymptotic rate with RVQ and an optimal receiver, given by \eqref{wt2}
and \eqref{wt1}. Hence $ \mathcal{R}_{\rvq;\mf}$ with $\hat{B} =
\infty$ is finite.  As for the analysis of the optimal receiver, this
discrepancy is again due to the fact that the cdf of $R_{\mf}^{N_r}$,
which has compact support, is being approximated by a Gaussian cdf
with infinite support, and also because the dependence among the sum
rates $R^{N_r}_{\mf} (\bm{V}_j)$ is being ignored.

%-------------------------
\subsection{MMSE receiver}
%-------------------------

Substituting \eqref{mmser} into \eqref{sinr} gives the SINR at the
output of MMSE receiver for the $k$th symbol stream
\begin{equation}
  \gamma_{k;\mmse} = \bm{v}_{k}^{\dag} \bm{H}^{\dag} \left( \sum_{i \ne
  k} \bm{H} \bm{v}_{i} \bm{v}_{i}^{\dag} \bm{H}^{\dag} + K \varn
  \bm{I}_{N_r} \right)^{-1} \bm{H} \bm{v}_{k} .
\end{equation}
As for the matched filter receiver, given
a codebook $\mathcal{V}$, we approximate the pdf of the
instantaneous sum rate
\begin{equation}
  R_{\mmse}^{N_r} = \frac{1}{N_r} \sum_{k = 1}^K \log( 1 + \gamma_{k;\mmse})
\end{equation}
as a Gaussian pdf with mean
\begin{equation}
  \mu_{\mmse} = \lim_{(N_t,N_r)\to\infty} R_{\mmse}^{N_r} = 
\frac{\K}{\N} \log ( 1 + \gamma_{\mmse} ) 
\end{equation}
where the large system SINR is given by \cite{tse00}
\begin{equation}
\begin{split}
 \gamma_{\mmse} &= \frac{1 - \K/\N}{2 \varn} - \frac{1}{2} \\
     &\quad + 
     \sqrt{\frac{(1 - \K/\N)^2}{4 \sigma_n^4} + 
     \frac{1 + \K/\N}{2 \varn} + \frac{1}{4}} .
\end{split}
\label{btf}
\end{equation}
As for the matched filter, the asymptotic variance
$\sigma^2_{\mmse}$ can be obtained via numerical simulation.
 
In analogy with \eqref{rvqmf}, the asymptotic rate with RVQ and the
MMSE receiver is given by
\begin{equation}
  \mathcal{R}_{\rvq;\mmse} \approx \tilde{\mathcal{R}}_{\rvq;\mmse} =
  \mu_{\mmse} + \sigma_{\mmse} \sqrt{2 \hat{B} \log 2} .
\end{equation}
As for the matched filter receiver, when $\hat{B}$ is large,
$\tilde{\mathcal{R}}_{\mmse}$ over-estimates $\mathcal{R}_{\mmse}$.
For $\hat{B} = \infty$, $\mathcal{R}_{\mmse} = \mathcal{R}_{\mf} =
\mathcal{I}_{\rvq}$ with the optimal receiver, given by \eqref{wt2}.

%-----------------------------
\subsection{Numerical Results}
%-----------------------------

\begin{figure}
  \centering
  \includegraphics[width=3.25in]{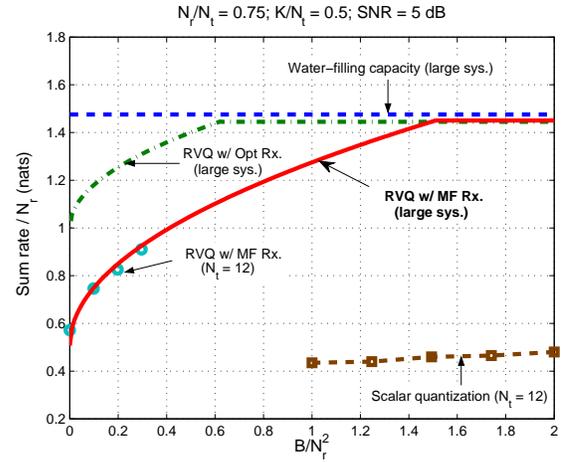}
  \caption{Sum rate per receive antenna versus normalized feedback
  bits with a matched filter receiver.  Results are shown for RVQ
  (asymptotic and $N_t = 12$) and scalar quantization.  Also shown are
  results for the optimal receiver, and the water-filling capacity
  with infinite feedback.}
  \label{fig_mf}
\end{figure}

Fig.~\ref{fig_mf} compares the approximation for asymptotic RVQ
performance with a matched filter receiver from \eqref{rvqmf} with
simulated results for $N_t = 12$, $\N = 0.75$, $K/N_t = 1/2$, and
$\text{SNR} = 5 \, \text{dB}$.  Also shown for comparison
are the asymptotic rate for RVQ with an optimal receiver, 
derived in Section~\ref{opt}, the water-filling capacity 
($\hat{B}= \infty$), and the rate achieved with a scalar quantizer
for each coefficient.\footnote{For the scalar quantization results
the available bits are spread evenly over the corresponding fraction of
precoding coefficients. The remaining coefficients are set to one.}
For the case shown, the analytical
approximation gives an accurate estimate of the performance of the
finite size system with limited feedback.  
The capacity with the water-filling power allocation is only slightly
greater than that achieved with the on-off power allocation.
The optimal receiver requires $\hat{B} \approx 0.6$ bit/dimension to
achieve the capacity corresponding to unlimited feedback \eqref{wt2},
whereas the matched filter requires $1.2$ feedback bits per dimension
to reach that capacity. 
For other target rates, these curves illustrate
the trade-off between feedback and receiver complexity.

\begin{figure}
  \centering
  \includegraphics[width=3.25in]{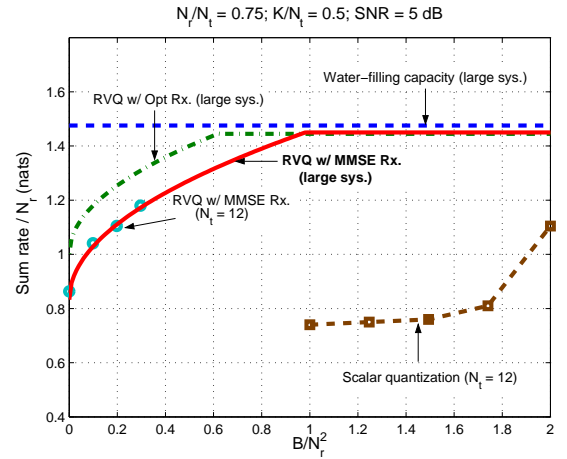}
  \caption{Sum rate per receive antenna versus normalized feedback
  bits with a linear MMSE receiver.  Results are shown for RVQ
  (asymptotic and $N_t = 12$) and scalar quantization.  Also shown are
  results for the optimal receiver, and the water-filling capacity
  with infinite feedback.}
  \label{fig_mmse}
\end{figure}

Fig.~\ref{fig_mmse} shows the same set of results
as those shown in Fig.~\ref{fig_mf}, but with an MMSE receiver.
These results show that for the parameters selected, the MMSE receiver
performs nearly as well as the optimal receiver, and requires
substantially less feedback than the matched filter to achieve
a target rate.  Again the asymptotic approximation
accurately predicts the performance 
of a system with a relatively small number of antennas.

%%%%%%%%%%%%%%%%%%%%%%%
\section{Conclusions}
\label{conclude}
%%%%%%%%%%%%%%%%%%%%%%%

We have studied the capacity of single-user MISO and MIMO fading
channels with limited feedback.  The feedback specifies a transmit
precoding matrix, which can be optimized for a given channel
realization.  We first considered the performance with a rank-one
precoding matrix (beamformer), and showed that the RVQ codebook is
asymptotically optimal. Exact expressions for the asymptotic mutual
information for MISO and MIMO channels were presented, and reveal how
much feedback is required to achieve a desired performance.  For the
cases considered, one feedback bit for each precoder coefficient can
achieve close to the water-filling capacity. Perhaps more important
than the increase in capacity provided by this feedback is the
associated simplification in the coding and decoding schemes that can
achieve a rate close to capacity.

The performance of a precoding matrix with rank $K > 1$ was also
evaluated with RVQ. Although numerical examples and our beamforming
results ($K=1$) suggest that RVQ is also asymptotically optimal in
this case, proving this is an open problem.  To compute the asymptotic
achievable rate for RVQ with both optimal and linear receivers, the
achievable rate with a random channel and fixed precoding matrix is
approximated as a Gaussian random variable.  The asymptotic rate then
depends on the asymptotic mean and variance of this random variable.
Although the asymptotic variance appears to be difficult to compute
analytically, it can be easily obtained by simulation.  Numerical
results have shown that the resulting approximation accurately
estimates the achievable rate with limited feedback for finite-size
systems of interest.

Numerical examples comparing the performance of optimal and linear
receivers have shown that the linear MMSE receiver requires little
additional feedback, relative to the optimal receiver, to achieve a
target rate close to the water-filling capacity.  The matched filter
requires significantly more feedback than the MMSE receiver (more than
0.5 bit per degree of freedom for the cases shown).  At low feedback
rates the achievable rate with RVQ is generally much greater than that
associated with scalar quantization.  Of course, this comes at a price
of high complexity, since the receiver is assumed to compute the
performance metric for every entry in the codebook. Other reduced
complexity schemes for quantizing a beamforming vector are presented
in \cite{isssta02,cdma04,Ryan07}.

Key assumptions for our results are that the channel is stationary and
known at the receiver, and that the channel elements are {\em i.i.d.}
Depending on user mobility and associated Doppler shifts, the channel
may change too fast to allow reliable channel estimation and
feedback. In that case, feedback of channel {\em statistics}, as
proposed in \cite{visotsky01,zhou02,simon03,jafar04,raghavan}, can
exploit correlation among channel elements.  The design of
quantization codebooks for precoders, which takes correlation into
account, is addressed in \cite{raghavan}.  The effect of channel
estimation error on the performance of limited feedback beamforming
with finite coherence time (i.e., block fading) is presented in
\cite{isit06,wcnc07}.

We have also assumed that the channel gains are not
frequency-selective.  Limited feedback schemes for frequency-selective
scalar channels are discussed in \cite{sun08}, and could be combined
with the quantization schemes considered here.  Finally, the approach
presented here for a single-user MIMO channel can also be applied to
multi-user models. Quantization of beamformers for the MIMO downlink
have been considered in \cite{sharif,jindal,goldsmith_jindal_isit}.
In that scenario the potential capacity gain due to feedback is
generally much more than for the single-user channel considered
here. The benefits of limited feedback for related models (e.g.,
frequency-selective MIMO downlink) are currently being studied.

%%%%%%%%%%%%%
\appendix
%%%%%%%%%%%%%

%%%%%%%%%%%%%%%%%%%%%%%%%%%%%%%%%%%%%%%%%%%%%
\subsection{Proof of Theorem~\ref{thm:miso}}
\label{proof_miso}
%%%%%%%%%%%%%%%%%%%%%%%%%%%%%%%%%%%%%%%%%%%%%

Given $\bm{h}$, the receiver selects the quantized
beamforming vector $\hat{\bv}$ to maximize the instantaneous rate
in \eqref{hbv}. Since $\log$ is monotonically increasing, the
quantized beamforming vector is given by 
\begin{equation}
   \hat{\bv} = \arg \max_{1 \le j \le 2^B} \left\{ Y_j =
      |\bv_j^{\dag}\bm{h}|^2 /\|\bm{h}\|^2 \right\} . 
\end{equation} 
Since the codebook entries $\bv_j$, $j = 1, \cdots, 2^B$, 
are {\em i.i.d.}, the $Y_j$'s, given $\bm{h}$, are also 
{\em i.i.d.} with the cdf \cite{chun_love}
\begin{equation}
\label{Fyy}
  F_{Y | \bm{h}} (y) = 1 - (1 - y)^{N_t - 1}, \quad 0 \le y \le 1 .
\end{equation}

We wish to determine the distribution of $\max_j Y_j$ given
$\bm{h}$. From \cite[Theorem 2.1.2]{galambos} it follows that
\begin{equation} 
  \frac{\max_j Y_j - a_n}{b_n} \stackrel{\mathcal{D}}{\longrightarrow} 
  \mathcal{Y}
\label{fmj}
\end{equation}
where
$\mathcal{Y} $ is a Weibull random variable having distribution
\begin{equation}
   H_{\gamma} (x) = \left\{
   \begin{array}{l@{\quad}l}
     1, & x \ge 0 \\
     \exp ( - (- x)^{\gamma}), & x < 0
  \end{array}
  \right. ,
\label{wb}
\end{equation}
$\mathcal{D}$ denotes convergence in distribution, and
$a_n$ and $b_n$ are normalizing sequences, where $n = 2^B$.
Specifically, the theorem requires that
$\omega(F_{Y | \bm{h}}) = \sup \{y: F_{Y | \bm{h}} (y) < 1 \}$ be
finite, and that the distribution function $F^{\ast}_{Y | \bm{h}}
(y) = F_{Y | \bm{h}} (
\omega(F_{Y | \bm{h}}) - 1/y ), y > 0$ satisfies, for all $y > 0$,
\begin{equation}
\label{lti}
  \lim_{t \to \infty} \frac{1 - F^{\ast}_{Y | \bm{h}} (t y)}{1 -
  F^{\ast}_{Y | \bm{h}}
  (t)} = y^{- \gamma}
\end{equation}
where the constant $\gamma > 0$.  

Substituting the expression for 
$F_{Y | \bm{h}}$ in \eqref{Fyy} into
\eqref{lti}, where $\omega(F_{Y | \bm{h}}) = 1$, gives
\begin{eqnarray}
   \lim_{t \to \infty} \frac{1 - F^{\ast}_{Y | \bm{h}} (t y)}{1 -
    F^{\ast}_{Y | \bm{h}}
    (t)} 
    & = & \lim_{t \to \infty} \frac{\left( 1/(ty)
      \right)^{N_t - 1}}{\left( 1/t \right)^{N_t - 1}} \\ 
    & = & y^{-(N_t - 1)}
\end{eqnarray}
so that \cite[Theorem 2.1.2]{galambos} applies when $N_t > 1$.  
Furthermore, the normalizing constants are given by
\begin{equation}
   a_n = \omega(F_{Y | \bm{h}}) = 1
\end{equation}
and
\begin{eqnarray}
   b_n & = & \omega(F_{Y | \bm{h}}) - \inf \left\{ 
  y : 1 - F_{Y | \bm{h}} (y) \le
  \frac{1}{n}  \right\} \\
     & =  & 1 - F^{-1}_{Y | \bm{h}} \left( 1 - \frac{1}{n} \right)
     =  \left( \frac{1}{n} \right)^{\frac{1}{N_t - 1}} .
\end{eqnarray}

To take the limit as $N_t \to \infty$,
we will assume that the channel vector $\bm{h}$ 
contains the first $N_t$ elements of an infinite-length
{\em i.i.d.} complex Gaussian vector $\bar{\bm{h}}$.
Rearranging terms in \eqref{fmj} and taking the large system limit gives
\begin{align}
  \lim_{(N_t,n) \to \infty} & E_{\mathcal{V}}[ \max_j Y_j | \bar{\bm{h}}] \\
    & =   \lim_{(N_t,n) \to \infty} a_n + b_n E [\mathcal{Y} ]\\
    & =  1 - \lim_{(N_t,n) \to \infty} \left( \frac{1}{n} 
        \right)^{\frac{1}{N_t - 1}} \Gamma \left( 1 - \frac{1}{N_t - 1} 
        \right) \\
    & =  1 - \lim_{(N_t,B) \to \infty} 2^{-\frac{B}{N_t - 1}} \\
    & =  1 - 2^{-\B}
  \end{align}
where the gamma function $\Gamma(z) = \int_0^{\infty} t^{z-1} \me^{-t} 
\, \diff t$ and we have used the fact that $E [\mathcal{Y} ] = 
-\Gamma(1 - 1/(N_t-1))$ \cite{castillo}.

From \eqref{fmj}, as $(n, N_t) \to \infty$,
\begin{equation}
  \var[ \max_j Y_j | \bar{\bm{h}} ] - b_n^2 \var[\mathcal{Y}] \to 0 .
\end{equation}
Since $\var[\mathcal{Y}] = \Gamma(1 - 2/(N_t-1)) - \Gamma^2(1 - 1/(N_t
-1))\to 0$, it follows that $\var[ \max_j Y_j | \bm{h} ] \to 0$.
This establishes that given $\bar{\bm{h}}$,
\begin{equation}
  \max_j Y_j \to 1 - 2^{-\B}
\end{equation}
in the mean square sense.
The asymptotic rate difference is given by
\begin{align}
  \mathcal{I}_{\rvq}^{\triangle} 
  & =  \lim_{(N_t,B) \to \infty}
\left( \log ( 1 + \rho \| \bm{h} \|^2
  \max_{1 \le j \le 2^B} Y_j ) \right) - \log ( \rho N_t ) \\ 
  & =  \lim_{(N_t,B) \to \infty}
\left( \log \left( \frac{1}{\rho N_t} +
  \frac{1}{N_t} \| \bm{h} \|^2
  \max_{1 \le j \le 2^B} Y_j \right) \right)\\ 
  & =  \log(1 - 2^{-\B})
\end{align}
in the mean square sense,
since $\| \bm{h} \|^2 / N_t \to 1 $ almost surely.

%%%%%%%%%%%%%%%%%%%%%%%%%%%%%%%%%%%%%%%%%%%%%
\subsection{Proof of Theorem~\ref{beam_opt}}
\label{proof_opt_miso}
%%%%%%%%%%%%%%%%%%%%%%%%%%%%%%%%%%%%%%%%%%%%%

The rate difference associated with codebook $\mathcal{V}_{N_t}$ is
\begin{align}
  I^{\triangle}_{\mathcal{V}_{N_t}} & = \max_{1 \le j \le 2^B} 
\log \left(
  \frac{1}{\rho N_t} + \frac{1}{N_t} | \bm{h}^{\dag} \bm{v}_j^{N_t}
  |^2 \right) \\ 
  & = \log \left( \frac{1}{\rho N_t} + \max_{1 \le j \le 2^B}
  \frac{1}{N_t} | \bm{h}^{\dag} \bm{v}_j^{N_t} |^2 \right) .
\end{align}
Taking expectation of the rate difference with respect
to $\bm{h}$ and applying Jensen's inequality, we obtain
\begin{align}
  E [ I^{\triangle}_{\mathcal{V}_{N_t}} ] & \le \log \left( \frac{1}{\rho
    N_t} + E [ \frac{1}{N_t} | \bm{h}^{\dag} \hat{\bm{v}}^{N_t} |^2]
  \right) \\ 
  & = \log \left( \frac{1}{\rho N_t} + E [ \frac{1}{N_t} \| \bm{h} \|^2 ] 
  E [ \mu ] \right)
\label{lfr}
\end{align}
where the optimal beamforming vector
\begin{equation}
  \hat{\bm{v}}^{N_t} = \arg \max_{1 \le j \le 2^B}  |
  \bm{h}^{\dag} \bm{v}_j^{N_t} |^2  ,
\end{equation}
$\mu = |\bm{h}^{\dag} \hat{\bm{v}}^{N_t} |^2 / \| \bm{h} \|^2$,
and \eqref{lfr} follows from the fact that $\| \bm{h} \|^2$ and $\mu$
are independent \cite{chun_love}.  

We now derive an upper bound for $E [ |\bm{h}^{\dag}
  \hat{\bm{v}}^{N_t} |^2 / \| \bm{h} \|^2 ]$ .  From (30) in
\cite{mukkavilli03} we have
\begin{equation}
\begin{split}
   \mathrm{Pr} \{
   |\bm{h}^{\dag} \hat{\bm{v}}^{N_t} |^2 >\gamma s ~|~ &\| \bm{h} \|^2 = \gamma
   \}  \\
   &\le \left\{
   \begin{array}{l@{\quad}l}
     1, & 0 \le s < s^*  \\
     2^B ( 1 - s )^{N_t -1} , & s^* \le s \le 1
  \end{array}
  \right.
\end{split}
\label{pha}
\end{equation}
where 
\begin{equation}
  s^* = 1 - 2^{-\frac{B}{N_t - 1}} .
\end{equation}
Since the right-hand side of \eqref{pha} is independent of $\gamma$,
averaging over $\gamma$ gives
\begin{equation}
   \mathrm{Pr} \{ \mu > s \}  \le \left\{
   \begin{array}{l@{\quad}l}
     1, & 0 \le s < s^*  \\
     2^B ( 1 - s )^{N_t -1} , & s^* \le s \le 1
  \end{array}
  \right. .
\label{prm}
\end{equation}

Integrating by parts, we have that
\begin{align}
  E [ \mu ] & = \int_0^1 \mathrm{Pr} \{ \mu
  > x \} \, \diff x \\
  & = \int_0^{s^*} \mathrm{Pr} \{ \mu
  > x \} \, \diff x + \int_{s^*}^1 \mathrm{Pr} \{ \mu
  > x \} \, \diff x . \label{prs}
\end{align}
Substituting \eqref{prm} into \eqref{prs} and evaluating both
integrals gives
\begin{equation}
  E [ \mu ] \le 1 - 2^{-\frac{B}{N_t-1}} +
  \frac{1}{N_t}2^{-\frac{B}{N_t-1}} .
\label{Emb}
\end{equation}

Substituting $E [\|\bm{h}\|^2 ]=N_t$ and \eqref{Emb} 
into \eqref{lfr} gives
\begin{equation}
  E [ I^{\triangle}_{\mathcal{V}_{N_t}} ] \le 
\log \left( 1 - 2^{-\frac{B}{N_t-1}} +
  \frac{1}{N_t}2^{-\frac{B}{N_t-1}} + \frac{1}{\rho N_t} \right)
\end{equation}
and taking the large system limit gives
\begin{equation}
  \lim_{(N_t,B) \to \infty} E [ I^{\triangle}_{\mathcal{V}_{N_t}} ] 
\le \log(1 - 2^{-\B}).
\end{equation}
Theorem~\ref{thm:miso} states that RVQ achieves this upper bound, and
therefore upper bounds the asymptotic rate difference corresponding
to any quantization scheme.

%%%%%%%%%%%%%%%%%%%%%%%%%%%%%%%%%%%%%%%
\subsection{Proof of Theorem~\ref{ua}}
\label{proof_mimo}
%%%%%%%%%%%%%%%%%%%%%%%%%%%%%%%%%%%%%%%

We first prove the theorem for $\N \ge 1$.  Let $z = F_{\gamma |
\bar{\bH}}^{-1} (1 - 2^{-B})$.  Rearranging \eqref{gir} gives
\begin{equation}
  \lim_{\substack{(N_t, N_r) \to \infty \\ z \to \srvq}} \left[ 1 -
  F_{\gamma | \bar{\bH}} (z) \right]^{\frac{1}{N_t}} = 2^{-\B} .
\label{lnr}
\end{equation}
Next, we derive upper and lower bounds for the left-hand side of
\eqref{lnr} and show that they are the same.  The derivation of the
upper bound is motivated by the evaluation of a similar bound for CDMA
signature optimization in \cite{dai_ciss}.  That is,
\begin{align}
  1 - F_{\gamma | \bar{\bH}} (z) & =  \Pr \left\{ \left. \gamma_j 
> z \right| \bar{\bH} \right\} \\
  & = \Pr \left\{ \bv_j^{\dag} \bU \bLd
         \bU^{\dag} \bv_j > z | \bLd, \bU \right\}\\
    & =  \Pr \left\{ \left. \frac{\bw_j^{\dag}\bU \bLd
         \bU^{\dag} \bw_j}{\bw_j^{\dag} \bw_j} > z \right| \bLd, \bU
   \right\}
\end{align}
where $\gamma_j = \frac{1}{N_t} \bv_j^{\dag} \bH^{\dag} \bH \bv_j$
and we have applied the singular value decomposition 
$\frac{1}{N_t} \bH^{\dag} \bH = \bU \bLd \bU^{\dag}$, 
where $\bU$ is an $N_t \times N_t$ unitary matrix,
$\bLd = \mathrm{diag} \{ \lambda_1, \cdots, \lambda_{N_t} \}$, and the
eigenvalues are ordered as $\lambda_1 \ge \lambda_2 \ge
\cdots \ge \lambda_{N_t}$.  
Also, $\bw_j$ is an $N_t \times 1$ vector with independent, circularly
symmetric, zero-mean and unit-variance Gaussian elements. Both
$\bv_j$ and $\bw_j/\| \bw_j \|$ are isotropically distributed, i.e.,
$\bU\bw_j/\| \bw_j \|$ and $\bw_j / \| \bw_j \|$ have 
the same distribution, so that
\begin{align}
  1 - F_{\gamma | \bar{\bH}} (z) & =  1 - F_{\gamma | \bLd} (z) \\
   & =  \Pr \left\{
         \left. \frac{\sum_{i=1}^{N_t} \lambda_i
         w_i^2}{\sum_{i=1}^{N_t} w_i^2} > z \right| \bLd \right\} \\
   & =  \Pr \left\{ \left. - \sum_{i=1}^{N_t} ( z - \lambda_i ) 
         w_i^2 > 0 \right| \bLd
         \right\} \\
   & =  \Pr \left\{ \left. - \rho \sum_{i=1}^{N_t} ( z - \lambda_i ) 
         w_i^2 >
         0 \right| \bLd \right\}, \quad \forall \rho > 0 \\
   & =  \Pr \left\{ \left. \exp \left( - \rho \sum_{i=1}^{N_t} 
         ( z - \lambda_i ) w_i^2 \right) > 1 \right| \bLd \right\}
\end{align}
where $\{ w_i \}$ are elements of $\bw_j$.  (We omit the index $j$ to
simplify the notation.)  Applying Markov's
inequality and the independence of the $w_i$'s gives
\begin{align}
  1 - F_{\gamma | \bLd} (z) & \le  E_{\{ w_i \}} \left[ \exp \left\{
         \left. - \rho \sum_{i=1}^{N_t} ( z - \lambda_i ) w_i^2
         \right\} \right| \bLd \right] \\
  & =  \prod_{i=1}^{N_t} E_{w_i} \left[
     \exp \left\{ - \rho ( z - \lambda_i ) w_i^2 \right\} | \lambda_i
     \right]\\ 
   & =  \prod_{i=1}^{N_t} \int^\infty_0 \exp
     \left\{ - \rho ( z - \lambda_i ) x \right\} \me^{-x} \diff x \\ 
   & =  \prod_{i=1}^{N_t} \int^\infty_0 \exp \left\{ -(1 + \rho( z -
     \lambda_i )) x \right\} \diff x \\ \label{piN}
  & =  \prod_{i=1}^{N_t} \frac{1}{1 + \rho ( z - \lambda_i )} \\
    & =  \exp \left\{ - \sum_{i = 1}^{N_t} \log(1 + \rho ( z - \lambda_i
      )) \right\}
\end{align}
when $ 1 + \rho(z - \lambda_i) > 0$ 
for all $i$, or $\rho < 1/(\lambda_1 - z)$.
Taking the large system limit, we obtain
\begin{equation}
  \lim_{\substack{(N_t, N_r) \to \infty \\ z \to \srvq}} \left[ 1 -
  F_{\gamma | \bLd} (z) \right]^{\frac{1}{N_t}} \le \exp \{ -
  \Phi(\srvq, \rho) \} 
\label{lsi}
\end{equation}
for $0 < \rho < \frac{1}{\lmax - \srvq}$, where
\begin{equation}
  \Phi(\srvq, \rho) \triangleq \int^b_a \log (1 + \rho (\srvq -
  \lambda)) g (\lambda) \diff \lambda ,
\label{Pst}
\end{equation}
$g( \lambda )$ is given by 
\eqref{eq:marcenko-pastur1}-\eqref{eq:marcenko-pastur2}, and
$\lmax = \lim_{(N_t,N_r) \to \infty} \lambda_1 = (1+ \sqrt{\N})^2$.  
To tighten the upper bound, we minimize
\eqref{lsi} with respect to $\rho$, i.e.,
\begin{equation}
  \lim_{\substack{(N_t, N_r) \to \infty \\ z \to \srvq}} \left[ 1 -
  F_{\gamma | \bLd} (z) \right]^{\frac{1}{N_t}} \le \exp \{ -
  \Phi(\srvq, \rho^*) \} \label{sNtr}
\end{equation}
where
\begin{equation}
  \rho^* = \arg \max_{0 < \rho < \frac{1}{\lmax
  - \srvq}} \Phi(\srvq, \rho) .
\label{eq:rhoopt}
\end{equation}
A similar expression for RVQ performance when used
to quantize signatures for CDMA is derived in
\cite{dai_ciss}.

To derive the lower bound, we use a change of measure. (A similar
approach was used in \cite[Section 1.2]{shwartz}.)  Let $y_i
\triangleq (\lambda_i - z)w_i^2$, which is a scaled exponential random
variable with cdf $F_i(\cdot)$.  We define the new distribution
\begin{equation}
  G_i(x) \triangleq \frac{1}{M_i(\rho^*)} \int_{-\infty}^x \me^{\rho^*
    y} \, \diff F_i(y) ,
\label{Gix}
\end{equation}
so that
\begin{equation}
  M_i(\rho^*) \diff G_i(x) = \me^{\rho^*x}\diff F_i(x),
\label{Mir}
\end{equation}
where $\rho^*$ is given in \eqref{eq:rhoopt}, and the moment 
generating function for $y_i$ is
\begin{align}
  M_i(\theta) & \triangleq E [ \me^{\theta y_i} ] \\
     & = \int_0^\infty \me^{\theta(\lambda_i - z) x} \me^{-x} \, \diff x \\
     & = \frac{1}{1 + \theta(z - \lambda_i)} .
\end{align}  

Applying the change of measure \eqref{Mir}, we have
\begin{align}
  &1 - F_{\gamma | \bLd} (z)  \nonumber \\
    & = \Pr \{ \sum_{i=1}^{N_t} y_i > 0\} \\ 
    & = \idotsint {\bf 1}[\sum_{i=1}^{N_t} y_i > 0] \, \diff F_1(y_1)
    \cdots \diff F_{N_t}(y_{N_t}) \\ 
    & = \idotsint {\bf
    1}[\sum_{i=1}^{N_t} y_i > 0] \me^{-\rho^* \sum_{i=1}^{N_t} y_i}
    \me^{\rho^*y_1}\diff F_1(y_1) \\
     &\qquad \cdots \me^{\rho^*y_{N_t}} \diff
    F_{N_t}(y_{N_t}) \nonumber \\ 
   & = \prod_{i=1}^{N_t} M_i(\rho^*) \idotsint
    {\bf 1}[\sum_{i=1}^{N_t} y_i > 0] \me^{-\rho^* \sum_{i=1}^{N_t}
    y_i} \,\diff G_1(y_1) \\
   &\qquad \cdots \diff G_{N_t}(y_{N_t}) \nonumber
\end{align}
where
\begin{equation}
  {\bf 1}[x > 0] = \left\{ 
  \begin{array}{r@{\quad:\quad}l}
    1 & x > 0\\
    0 & x \le 0
  \end{array}\right. .
\end{equation}
For any $\epsilon >0$,
\begin{align}
  &1 - F_{\gamma | \bLd} (z) \nonumber \\
& \ge \prod_{i=1}^{N_t} M_i(\rho^*) \idotsint 
      {\bf 1}[\epsilon N_t \ge \sum_{i=1}^{N_t} y_i > 0]\me^{-\rho^* \sum_{i=1}^{N_t} y_i} 
      \,\\
  & \qquad \diff G_1(y_1) \cdots \diff G_{N_t}(y_{N_t}) \nonumber\\
  & \ge \prod_{i=1}^{N_t} M_i(\rho^*) \me^{-\rho^* \epsilon N_t} \idotsint 
      {\bf 1}[\epsilon N_t \ge \sum_{i=1}^{N_t} y_i > 0] 
      \,\\
  &\qquad \diff G_1(y_1) \cdots \diff G_{N_t}(y_{N_t}) \nonumber\\
  & = \prod_{i=1}^{N_t} M_i(\rho^*)  \me^{-\rho^* \epsilon N_t} 
     \Pr\{ \epsilon N_t \ge \sum_{i=1}^{N_t} \tilde{y}_i > 0 \} 
\label{yi0}
\end{align}
where the $\tilde{y}_i$'s are independent random variables with cdf
$G_i(\cdot)$, and the second inequality follows since $\rho^* > 0$.
To determine the probability on the right-hand side of \eqref{yi0}, we
first compute the mean of $\tilde{y}_i$,
\begin{align}
  m_i &= \int y \, \diff G_i(y) \\
       & = \frac{1}{M_i(\rho^*)} \int y \me^{\rho^* y} \diff F_i(y) \\
       & = (1 + \rho^*(z - \lambda_i)) \int_0^\infty (\lambda_i - z) x 
         \me^{\rho^* (\lambda_i - z) x} \me^{-x}\, \diff x \\
       & = \frac{\lambda_i - z}{1 + \rho^*(z - \lambda_i)} .
\end{align}
Therefore
\begin{equation}
  \frac{1}{N_t} \sum_{i=1}^{N_t} m_i = \frac{1}{N_t} \sum_{i=1}^{N_t} 
      \frac{\lambda_i - z}{1 + \rho^*(z - \lambda_i)}
\end{equation}
and the asymptotic mean
\begin{align}
  m_{\infty} & = \lim_{\substack{(N_t, N_r) \to \infty \\ z \to \srvq}}
   \frac{1}{N_t} \sum_{i=1}^{N_t} m_i\\
   & = \int_a^b 
    \frac{\lambda - \srvq}{1 +
    \rho^*(\srvq - \lambda)} g(\lambda) \, \diff \lambda 
     < \infty  .\label{mean0}
\end{align}
Similarly, since $\tilde{y}_i$ is exponentially distributed,
the variance of $\tilde{y}_i$ is
\begin{equation}
  \sigma_i^2 = \left( \frac{\lambda_i - z}{1 + \rho^*(z - \lambda_i)}
  \right)^2 < \infty
\end{equation}
and the asymptotic variance
\begin{align}
  \sigma_{\infty}^2 & =  \lim_{\substack{(N_t, N_r) \to \infty \\ z \to \srvq}}
   \frac{1}{N_t} \sum_{i=1}^{N_t} \sigma_i^2\\
  & = \int_a^b \left( \frac{\lambda
    - \srvq}{1 + \rho^*(\srvq - \lambda)} \right)^2 g(\lambda) \,
  \diff \lambda < \infty .
\end{align}
Both the asymptotic mean and variance are finite.

Since the $\tilde{y}_i$'s are independent with finite mean and variance,
the central limit theorem implies that the cdf for
\begin{equation}
  T \triangleq \frac{\sum_{i=1}^{N_t} \tilde{y}_i - \sum_{i=1}^{N_t}
    m_i}{\sqrt{\sum_{i=1}^{N_t} \sigma^2_i}}
\end{equation}
converges to a Gaussian cdf with zero mean and unit variance.  
Therefore we have
\begin{align}
   &\Pr\{ 0 < \sum_{i=1}^{N_t} \tilde{y}_i \le \epsilon N_t \} \nonumber\\
   & = \Pr
   \{-\sqrt{N_t} a_{N_t} < T \le -\sqrt{N_t} a_{N_t} + \sqrt{N_t}
   \epsilon \frac{1}{b_{N_t}} \} \\ 
    & = F_T(-\sqrt{N_t} a_{N_t} +
   \sqrt{N_t} \epsilon \frac{1}{b_{N_t}}) - F_T(-\sqrt{N_t} a_{N_t})
   \label{sNt}
\end{align}
where $F_T(\cdot)$ is the cdf for $T$ and 
\begin{gather}
  a_{N_t} \triangleq \frac{\sum_{i=1}^{N_t} m_i
    /N_t}{\sqrt{\sum_{i=1}^{N_t} \sigma_i^2 / N_t}} \to
  \frac{m_{\infty}}{\sigma_{\infty}} ,\\ 
  b_{N_t} \triangleq
  \sqrt{\sum_{i=1}^{N_t} \sigma_i^2 / N_t} \to \sigma_{\infty} .
\end{gather}

Let $\phi(\cdot)$ denote a Gaussian cdf with zero mean and unit
variance.  We can rewrite \eqref{sNt} as
\begin{equation}
\begin{split}
  \Pr\{ 0 < \sum_{i=1}^{N_t} \tilde{y}_i \le \epsilon N_t \} =
  \phi(-\sqrt{N_t} a_{N_t} + \sqrt{N_t} \epsilon \frac{1}{b_{N_t}}) \\
  -
  \phi(-\sqrt{N_t} a_{N_t}) + \zeta_{N_t} - \xi_{N_t}
\end{split}
\label{PrN}
\end{equation}
where
\begin{align}
  \zeta_{N_t} & \triangleq F_T(-\sqrt{N_t} a_{N_t} + \sqrt{N_t} \epsilon
  \frac{1}{b_{N_t}}) \nonumber\\
  &\quad - \phi(-\sqrt{N_t} a_{N_t} + \sqrt{N_t} \epsilon
  \frac{1}{b_{N_t}}), \\
  \xi_{N_t} & \triangleq F_T(-\sqrt{N_t} a_{N_t}) - \phi(-\sqrt{N_t} a_{N_t}) .
\end{align}

Applying the Berry-Ess\'{e}en theorem \cite{HBNormal}, 
we can bound both $\zeta_{N_t}$ and $\xi_{N_t}$ for large $N_t$ as 
\begin{equation}
  |\zeta_{N_t}|, \quad  |\xi_{N_t}| \le \frac{C}{\sqrt{N_t}}
\label{zqe}
\end{equation}
where $C$ is a positive constant that depends
on the variance and third moment of $\tilde{y}$.  
Similar to the mean and variance, we can
show that the third moment is also finite.

We can now evaluate
\begin{align}
&\phi(-\sqrt{N_t} a_{N_t} + \sqrt{N_t} \epsilon \frac{1}{b_{N_t}}) -
\phi(-\sqrt{N_t} a_{N_t}) \nonumber\\
& = \frac{1}{\sqrt{2\pi}}\int_{-\sqrt{N_t}
  a_{N_t}}^{-\sqrt{N_t} a_{N_t} + \sqrt{N_t} \epsilon
  \frac{1}{b_{N_t}}} \me^{-t^2/2}\, \diff t\\ & \le
\frac{1}{\sqrt{2\pi}} \me^{-N_t(a_{n_t} -
  \epsilon/b_{N_t})^2}\sqrt{N_t}\epsilon \frac{1}{b_{N_t}} .
\label{ebn}
\end{align}
Substituting \eqref{zqe} and \eqref{ebn} into \eqref{PrN}, we have
\begin{equation}
  \Pr\{ 0 < \sum_{i=1}^{N_t} \tilde{y}_i \le \epsilon N_t \} = O(1/\sqrt{N_t})
\end{equation}

Taking the large system limit and applying L'Hopital's rule,
it follows that
\begin{equation}
   \lim_{\substack{(N_t, N_r) \to \infty \\ z \to \srvq}} [\Pr\{ 0 <
     \sum_{i=1}^{N_t} \tilde{y}_i \le \epsilon N_t \}]^{\frac{1}{N_t}}
   = 1. \label{l1l}
\end{equation}
Taking the $N_t$th root and large system limit on both sides of
\eqref{yi0} gives
\begin{equation}
   \lim_{\substack{(N_t, N_r) \to \infty \\ z \to \srvq}} \left[ 1 -
  F_{\gamma | \bLd} (z) \right]^{\frac{1}{N_t}}  \ge 
    \exp\{-\Phi(\srvq, \rho^*)\} \label{gm0}
\end{equation}
where we use \eqref{l1l} and let $\epsilon \to 0$.

The lower bound in \eqref{gm0} is exactly the 
upper bound \eqref{sNtr}. Therefore,
\begin{align}
  \lim_{\substack{(N_t, N_r) \to \infty \\ z \to \srvq}} \left[ 1 -
  F_{\gamma | \bLd} (z) \right]^{\frac{1}{N_t}} & =  \exp \{ -
  \Phi(\srvq, \rho^*) \} \\
    & =  2^{-\B}
\end{align}
and the asymptotic RVQ received power satisfies the
fixed-point equation
\begin{equation}
  \Phi(\srvq, \rho^*) = \B \log(2)
\label{Psr}
\end{equation}
where $\rho^*$ is given by \eqref{eq:rhoopt}.
The goal of the rest of the proof is to simplify \eqref{Psr}.  

To determine $\rho^*$, we first compute
\begin{align}
  \frac{\partial \Phi (\srvq, \rho)}{\partial \rho} & =  \int_a^b \left[
        \frac{\srvq - \lambda}{1 + (\srvq - \lambda)\rho} \right] g(\lambda) 
        \, \diff \lambda \\
  & =  \frac{1}{\rho} + \frac{1}{\rho^2} \int_a^b
        \frac{1}{\lambda - \underbrace{\left(\frac{1}{\rho} + \srvq
       \right)}_y}
        g(\lambda) \, \diff \lambda \\
  & =  \frac{1}{\rho} + \frac{1}{\rho^2} \mathcal{S}_{\bLd} (y)
\label{f1r}
\end{align}
where $\mathcal{S}_{\bLd} (\cdot)$ is the Stieltj\'{e}s Transform of the
asymptotic eigenvalue distribution of $\bLd$.
Setting the derivative to zero and solving for $\rho$ gives
\begin{equation}
  \mathcal{S}_{\bLd} ( y ) = - \rho = \frac{1}{\srvq - y}
\label{mSy}
\end{equation}
as the only valid solution.  Substituting the expression for
$\mathcal{S}_{\bLd}$ given in \cite{marcenko} into \eqref{mSy} gives
\begin{equation}
\begin{split}
  \frac{(-1 + \N - y) \pm \sqrt{y^2 - 2 (\N + 1) y + (\N - 1)^2}}{2 y}\\
  = \frac{1}{\srvq - y} ,
\end{split}
\end{equation}
which simplifies to the quadratic equation
\begin{equation}
  (\N - \srvq) y^2 + [\srvq + \N \srvq + (\srvq)^2]y = 0 .
\end{equation}
Solving for $y$ gives $y = 0$ or $y = \srvq[1 + 1/(\srvq - \N)]$,
or equivalently, $\rho = -1/\srvq$ or $\rho = (\srvq - \N)/\srvq$.  
Since $\rho > 0$, we must have
\begin{equation}
  \rho^* = \frac{\srvq - \N}{\srvq} .
\label{rfs}
\end{equation}
Since
\begin{align}
  \frac{\partial^2 \Phi (\srvq, \rho)}{\partial \rho^2} 
       & = - \int_a^b \left( \frac{\lambda
    - \srvq}{1 + \rho^*(\srvq - \lambda)} \right)^2 g(\lambda) \,
  \diff \lambda \\
       & < 0,
\end{align}
therefore $\rho^*$ achieves a maximum.

By also evaluating $\Phi(\srvq, \rho)$ at the boundary points $\rho =
0$ and $\rho = 1/(\lmax - \srvq)$, we have
\begin{equation}
\begin{split}
&  \rho^* \\
& = \left\{ \begin{array}{l@{,\quad}l}
            \frac{\srvq - \N}{\srvq} & \N \le \srvq \le \N + \sqrt{\N}
            \\
            \frac{1}{(1 + \sqrt{\N})^2 - \srvq} & \N + \sqrt{\N} \le
            \srvq < (1 + \sqrt{\N})^2
           \end{array}
           \right. .
\end{split}
\end{equation}

To evaluate $\Phi( \srvq, \rho^* )$, we re-write \eqref{Pst} as
\begin{align}
  &\Phi( \srvq, \rho^* ) \nonumber\\
& =  \int_a^b \left[ \log(\rho^*) +
  \log \left( \left(\frac{1}{\rho^*} + \srvq \right) - \lambda \right)
  \right] g(\lambda) \, \diff \lambda\\
 & =  \log(\rho^*) + \int_a^b \log \left( \left(\frac{1}{\rho^*} 
    + \srvq \right) - \lambda \right) g(\lambda) \, \diff \lambda .
  \label{llh}
\end{align}
To evaluate the integral in \eqref{llh}, we apply the following Lemma.
\begin{lemma}
  For $x \ge (1 + \sqrt{\N})^2$,
  \begin{align}
    \Theta(x) & \triangleq  \int^b_a \log (x - \lambda) g(\lambda) \,
       \diff \lambda \\
     & =  \log(w(x)) + \sqrt{\N} u(x) - (\N-1)\log 
          \left(1 + \frac{u(x)}{\sqrt{\N}} \right) \label{Txw}
  \end{align}
where 
\begin{align}
  w(x) & =  \frac{(x - 1 - \N) + \sqrt{(x-1-\N)^2 - 4\N}}{2}, \\
  u(x) & =  \frac{(x - 1 - \N) - \sqrt{(x-1-\N)^2 - 4\N}}{2
    \sqrt{\N}} \label{uxf}.
\end{align}
\end{lemma}
The proof of this Lemma is similar to that given in \cite{rapajic00}
and is therefore omitted here.

For $\N + \sqrt{\N} \le \srvq < (1 + \sqrt{\N})^2$, we substitute
  $\rho^* = [(1 + \sqrt{\N})^2 - \srvq]^{-1}$ into \eqref{llh} to obtain
\begin{align}
  & \Phi( \srvq, [(1 + \sqrt{\N})^2 - \srvq]^{-1}) \\
  & =  - \log[(1 +
  \sqrt{\N})^2 - \srvq ] + \Theta \left( (1 +
  \sqrt{\N})^2 \right) \\
  & = - \log[(1 + \sqrt{\N})^2 - \srvq ] + \frac{1}{2}\N \log(\N) \nonumber\\
   &\quad -(\N - 1) \log ( 1 + \sqrt{\N}) +\sqrt{\N}\\
  & = \B \log(2) .
\end{align}
Solving for $\srvq$ gives \eqref{g1N}. 
Taking $\srvq = \N + \sqrt{\N}$ and solving for
$\B$ gives $\B^*$ in \eqref{bst}.

For $\N \le \srvq < \N + \sqrt{\N}$, or $0 \le \B \le \B^*$, 
we substitute 
$\rho^* = \frac{\srvq - \N}{\srvq}$ into \eqref{llh} to obtain
\begin{equation}
\begin{split}
  \Phi \left( \srvq, \frac{\srvq - \N}{\srvq} \right)
     =   \log(\srvq - \N) - \log( \srvq) \\
   + \Theta \left( \srvq +
     \frac{\srvq}{\srvq - \N} \right).
\end{split}
\label{psq}
 \end{equation}
To simplify \eqref{psq}, we let $\psi \triangleq \srvq - \N$
and re-write \eqref{psq} as
\begin{equation}
\begin{split}
  \Phi \left( \psi - \N, \frac{\psi}{\psi + \N}  \right)
     =   \log(\psi) - \log( \psi + \N) \\
  + \Theta \left( 1 + \N + \psi +
     \frac{\N}{\psi} \right).
\end{split}
\label{pNp}
 \end{equation}
After some manipulation we have
\begin{eqnarray}
  w \left( 1 + \N + \psi + \frac{\N}{\psi} \right) 
       & = &  \frac{\N}{\psi}, \\
  u \left( 1 + \N + \psi + \frac{\N}{\psi} \right)
       & = &  \frac{\psi}{\sqrt{\N}},
\end{eqnarray}
and
\begin{equation}
\begin{split}
  \Theta \left( 1 + \N + \psi + \frac{\N}{\psi} \right)
     = \log(\N) - \log(\psi) \\
   - (\N - 1)\log \left( 1 + \frac{\psi}{\N}
     \right) + \psi .
\end{split}
\label{Tl1}
\end{equation}
Substituting \eqref{Tl1} into \eqref{pNp}, we obtain
\begin{align}
  &\Phi \left( \psi - \N, \frac{\psi}{\psi + \N}  \right) \nonumber\\ 
       & =  \psi  - \N \log \left( 1 + \frac{\psi}{\N} \right)\\
   & =  \srvq - \N - \N \log ( \srvq ) + \N \log (\N). \label{sNN}
\end{align}
Setting this to $\B \log(2)$ and simplifying gives \eqref{lsr}.

For $\N < 1$, the asymptotic eigenvalue density of $\frac{1}{N_t}
\bH^{\dag} \bH$ is given by
\begin{equation}
  g(\lambda) = (1 - \N) \delta (\lambda) + 
      \frac{\sqrt{(\lambda - a)(b - \lambda)}}{2 \pi \lambda} .
\end{equation}
where $a$ and $b$ are given by 
\eqref{eq:marcenko-pastur1}-\eqref{eq:marcenko-pastur2}.
Following the same steps again
from \eqref{Pst} gives \eqref{g1N} and \eqref{lsr}.  This completes
the proof of Theorem~\ref{ua}.

%%%%%%%%%%%%%%%%%%%%%%%%%%%%%%%%%%%%%%%%%%%%%%%%%%%
\subsection{Derivation of \eqref{mu}-\eqref{ll1}}
\label{lemma_gauss}
%%%%%%%%%%%%%%%%%%%%%%%%%%%%%%%%%%%%%%%%%%%%%%%%%%%

To compute $ \mu_J $,  we first write
\begin{equation}
  J_j^{N_r} = \frac{1}{N_r} \sum_{k = 1}^{N_r} \log \left( 1 + \rho
  \frac{\N}{\K} \upsilon_k \right)
\end{equation}
where $ \upsilon_k $ is the $k$th eigenvalue of $ \bm{\Upsilon} =
\frac{1}{N_r} \bm{H} \bm{V}_j \bm{V}_j^{\dag} \bm{H}^{\dag} $. As 
$ (N_t, N_r, K) \to \infty $, the empirical eigenvalue distribution
converges to a deterministic function $ F_{\bm{\Upsilon}} (t) $.  The
asymptotic mean is given by
\begin{equation}
  \mu_J = \lim_{(N_t,N_r,K) \to \infty} E [ J_j^{N_r} ] =
  \int_0^{\infty} \log \left( 1 + \rho \frac{\N}{\K} t \right) \,
  \diff F_{\bm{\Upsilon}} (t) .
\label{jnm}
\end{equation}
A similar integral has been evaluated in \cite[Eq.~(6)]{rapajic00},
and the result can be directly applied to \eqref{jnm}, giving
\eqref{mu}.  

To compute the variance, we express $J_j^{N_r}$ differently 
by first performing the singular value decomposition 
$\bm{H} = \bm{V}_{\bm{H}}
\bm{\Sigma}_{\bm{H}} \bm{U}_{\bm{H}}^{\dag}$, where $\bm{V}_{\bm{H}}$
is the $N_r \times N_r$ left singular matrix, $\bm{U}_{\bm{H}}$ is the
$N_t \times N_r$ right singular matrix, and $\bm{\Sigma}_{\bm{H}}$ is
an $N_r \times N_r$ diagonal matrix.  Here we assume that $N_t \ge
N_r$. (The result for $N_t < N_r$ can be shown by a similar approach.)
We therefore have
\begin{align}
  J_j^{N_r} & =  \frac{1}{N_r} \log \det \left( \bm{I}_{N_r} + 
	\rho \bm{\Lambda} \bm{L}_j \right) \\
   & =  \frac{1}{N_r} \sum_{i = 1}^{N_r} \log 
      \left( 1 + \rho \eta_i \right) \label{f1nr}
\end{align}
where $\bm{\Lambda} = \frac{1}{K} \bm{\Sigma}^2_{\bH}$, $\bm{L}_j =
\bU^{\dag}_{\bH} \bV_j \bV_j^{\dag} \bU_{\bH}$, and $\eta_i$ is the
$i$th eigenvalue of $\bm{\Lambda} \bm{L}_j$.  To compute $\var
[J_j^{N_r} ]$, correlations between pairs of $\eta_i$'s are needed.
Although the joint distribution of eigenvalues is known, it is
complicated, so that computing the variance appears intractable.

To approximate the variance of $J_j^{N_r}$, we
substitute a Taylor series expansion for $\log(1 + \delta x)$
into \eqref{f1nr} to write
\begin{align}
  J_j^{N_r} & =  \frac{\rho}{N_r} \sum_{i=1}^{N_r} \eta_i -
  \frac{\rho^2}{2 N_r} \sum_{i=1}^{N_r} \eta_i^2 + \frac{\rho^3}{3 N_r}
  \sum_{i=1}^{N_r} \eta_i^3 + \ldots \\
  & =  \frac{\rho}{N_r} \tr\{ \bm{\Lambda} \bm{L} \} -
  \frac{\rho^2}{2 N_r} \tr\{ (\bm{\Lambda} \bm{L})^2 \} +
  \frac{\rho^3}{3 N_r} \tr\{ (\bm{\Lambda} \bm{L})^3 \} \nonumber\\
   & \quad + \ldots
\end{align}
for $\rho \eta_{\max} < 1$, where $\eta_{\max} = \max_i \eta_i$, and
is the maximum eigenvalue of $\bH \bV \bV^\dag \bH^\dag / K$.  Since
$\bH \bV$ is $N_r \times K$ and {\em i.i.d.}, $\eta_{\max}$ has
asymptotic value $(1+\sqrt{\N/\K})^2$.  If $\bar{K}/\bar{N_r} = 1$,
then the condition asymptotically becomes $\rho < 1/4$ (-6 dB).
Ignoring the terms of order $\rho^3$ and higher, we can approximate
the variance of $J_j^{N_r}$ at low SNR as
\begin{equation}
  \var [ J_j^{N_r} ] \approx \rho^2 \var \left[ \frac{1}{N_r} \tr\{
  \bm{\Lambda} \bm{L} \} \right] .
\label{vjnr}
\end{equation}
Letting $\bm{\Lambda} = \mathrm{diag} \{ \lambda_i \}$
and $l_{ij}$ denote the $(i,j)$th element of $\bm{L}$,
the first term in \eqref{vjnr} can be expanded as
\begin{equation}
\begin{split}
  \var [ \tr \{ \bm{\Lambda} \bm{L} \} | \bm{\Lambda} ] =
  \sum_{i=1}^{N_r} \lambda_i^2 \left( E [l_{ii}^2] - E^2 [ l_{ii} ]
  \right) \\
  + \sum_{i \ne j} \lambda_i \lambda_j \left( E [l_{ii}
  l_{jj}] - E [l_{ii}] E [l_{jj}] \right) .
\end{split}
\label{vtrbm}
\end{equation}

For a given $\bm{U}_{\bm{H}}$ and
random unitary $\bm{V}$ with $K = N_r$, Theorem~3 in \cite{roh_sp06}
states that $\bm{L}$ has a multivariate beta distribution with
parameters $N_r$ and $N_t - N_r$.  (The distribution of $\bm{L}$ is
not known for general $K$.)
From Theorem 2 in \cite{khatri}, we have
\begin{align}
  E [ l_{ii} ] &= \frac{N_r + 1}{N_t + 2} \label{elii} \\
  E [ l_{ii}^2 ] &= \frac{(N_r + 1)(N_r + 3)}{(N_t + 2)(N_t + 4)} \\
  E [l_{ii} l_{jj}] &= \frac{N_r(N_r+1)(N_t+4) +
  (N_r+1)(N_t-N_r+1)}{(N_t+1)(N_t+2)(N_t+4)},\nonumber\\
  & \quad ~~~ i \ne j, \label{eliiljj}
\end{align}
for $1\leq i,j \leq N_r$.
Substituting \eqref{elii}-\eqref{eliiljj} into \eqref{vtrbm} gives
\begin{equation} \begin{split}
  \var [ \tr \{ \bm{\Lambda} \bm{L} \} | \bm{\Lambda} ] =
  \left( \frac{1}{N_r} \sum_{i=1}^{N_r} \lambda_i^2 \right) \left(
  \N^2(1 - \N) + O\left( \frac{1}{N_r} \right) \right) \\
  + \left(
  \frac{1}{N_r^2} \sum_{i \ne j} \lambda_i \lambda_j \right) \left(
  (\N - 1)\N^3 + O\left( \frac{1}{N_r} \right) \right) .
\end{split} \end{equation}
Taking expectation with respect to $\bm{\Lambda}$, and the large
system limit, we have
\begin{equation}
  E_{\bm{\Lambda}} \left( 
\var [ \tr \{ \bm{\Lambda} \bm{L} \} | \bm{\Lambda} ] 
\right) \to 1 - \N .
\label{ebmv}
\end{equation}
Also, in the large system limit
\begin{eqnarray}
  \frac{1}{N_r} \sum_{i=1}^{N_r} \lambda_i^2 & \to & \int t^2 \diff
  F_{\bm{\Lambda}}(t) = \frac{1}{\N} \left( 1 +  \frac{1}{\N} \right) \\
  \frac{1}{N_r^2} \sum_{i \ne j} \lambda_i \lambda_j & \to &
  \left[ \int t \diff F_{\bm{\Lambda}}(t) \right]^2 = \frac{1}{\N^2}
\end{eqnarray}
where $F_{\bm{\Lambda}} (t)$ is the asymptotic distribution for the
diagonal elements of $\bm{\Lambda}$ or, equivalently, the asymptotic
eigenvalue distribution of $\bm{H} \bm{H}^{\dag}/N_r$.

Substituting \eqref{ebmv} into \eqref{vjnr}, we have
\begin{eqnarray}
  \sigma^2_J & = & \lim_{(N_r, N_t) \to \infty} N_r^2 \var [ J_j^{N_r} ]
  \\
  & \approx &  \rho^2 ( 1 - \N ) .
\end{eqnarray}

%%%%%%%%%%%%%%%%%%%%%%%%%%%
\section*{Acknowledgement}
%%%%%%%%%%%%%%%%%%%%%%%%%%%

The authors thank the anonymous reviewers for their detailed comments
and for pointing out mistakes in the proofs of Theorems~\ref{beam_opt}
and~\ref{ua}, which appeared in an earlier draft.

%%%%%%%%%%%%%%%%%%%%%%
% REFERENCE
%%%%%%%%%%%%%%%%%%%%%%

%%%% BIO'S %%%%%%%%%%%%%%%%%%%%%%%%%%

\begin{IEEEbiographynophoto}{Wiroonsak Santipach}
(S'00-M'06) received the B.S. ({\em summa cum laude}), M.S., and
Ph.D. degrees all in electrical engineering from Northwestern
University, Illinois, USA in 2000, 2001, and 2006, respectively.

He is currently a lecturer at the Department of Electrical
Engineering, Faculty of Engineering, Kasetsart University in Bangkok,
Thailand.  His research interests are in wireless communications, and
include performance evaluation of CDMA and MIMO system.
\end{IEEEbiographynophoto}

\begin{IEEEbiographynophoto}{Michael L. Honig}
(S'80-M'81-SM'92-F'97) received the B.S. degree in electrical
  engineering from Stanford University in 1977, and the M.S. and
  Ph.D. degrees in electrical engineering from the University of
  California, Berkeley, in 1978 and 1981, respectively. He
  subsequently joined Bell Laboratories in Holmdel, NJ, where he
  worked on local area networks and voiceband data transmission. In
  1983 he joined the Systems Principles Research Division at Bellcore,
  where he worked on Digital Subscriber Lines and wireless
  communications.  Since the Fall of 1994, he has been with
  Northwestern University where he is a Professor in the Electrical
  Engineering and Computer Science Department.  He has held visiting
  scholar positions at the Technical University of Munich, Princeton
  University, the University of California, Berkeley, Naval Research
  Laboratory (San Diego), and the University of Sydney.  He has also
  worked as a free-lance trombonist.

Dr. Honig has served as an editor for the IEEE Transactions on
Information Theory (1998-2000), the IEEE Transactions on
Communications (1990-1995), and was a guest editor for the European
Transactions on Telecommunications and Wireless Personal
Communications. He has also served as a member of the Digital Signal
Processing Technical Committee for the IEEE Signal Processing Society,
and as a member of the Board of Governors for the Information Theory
Society (1997-2002).  He is the recipient of a Humboldt Research Award
for Senior U.S. Scientists, and the co-recipient of the 2002 IEEE
Communications Society and Information Theory Society Joint Paper
Award.
\end{IEEEbiographynophoto}

\end{document}